# Sigmoid-Based Refined Composite Multiscale Fuzzy Entropy and t-Distributed Stochastic Neighbor Embedding Based Fault Diagnosis of Rolling Bearing

Zhanwei Jiang, Jinde Zheng*, Haiyang Pan, Ziwei Pan

School of Mechanical Engineering, Anhui University of Technology, Maanshan, China, 243032

**Abstract**： Multiscale fuzzy entropy (MFE) has been a prevalent tool to quantify the complexity of time series. However, it is extremely sensitive to the predetermined parameters and length of time series and it may yield an inaccurate estimation of entropy or cause undefined entropy when the length of time series is too short. In this paper the Sigmoid-based refined composite multiscale fuzzy entropy (SRCMFE) is introduced to improve the robustness of complexity measurement of MFE for short time series analysis. Also SRCMFE is used to quantify the dynamical properties of mechanical vibration signals and based on that a new rolling bearing fault diagnosis approach is proposed by combining SRCMFE with t-distributed stochastic neighbor embedding (t-SNE) for feature dimension and variable predictive models based class discrimination (VPMCD) for mode classification. In the proposed method, SRCMFE firstly is employed to extract the complexity characteristic from vibration signals of rolling bearing and t-SNE for feature dimension reduction is utilized to obtain a low dimensional manifold characteristic. Then VPMCD is employed to construct a multi-fault classifier to fulfill an automatic fault diagnosis. Finally, the proposed approach is applied to experimental data of rolling bearing and the results indicate that the proposed method can effectively distinguish different fault categories of rolling bearings.

## 1 Introduction

Rolling bearing plays an important part in rotating machines. In case the rolling bearing happens to failure, which will lead to the entire mechanical system breakdown and even accidents. Therefore, the health monitoring and fault detection of rolling bearing have been the focus of research and have been appealing to more and more researchers continuing to study. Vibration signal analysis methods have been the most significant methodology in the field of machinery fault diagnosis. However, many fault diagnosis methods of rolling bearings are often based on the assumption that the vibration signals are linear and stationary. However, the vibration signals often show nonlinear and non-stationary characteristic for the factors such as instantaneous variations in friction, damping or loading conditions and thus traditional linear signal analysis methods will no longer be valid.

* Corresponding author. Tel.:+86 18395585081.
*E-mail addresses*: jzwptt666888@gmail.com (Z. Jiang), lqdlzheng@126.com (J. Zheng).

With the development of nonlinear dynamics, lots of nonlinear dynamic parameters have been successfully applied in mechanical fault diagnosis. For example, approximate entropy (ApEn) proposed by Pincus [1] was used as a diagnostic tool for machine health monitoring [2]. However, ApEn algorithm depends heavily on the length of time series and the estimated value usually is smaller than the expected ones while processing dataset is too short. Sample entropy (SampEn) was proposed by Richman et al. [3] to overcome the shortcomings of ApEn. But the long-term structures in the multiple time scales fail to be captured by SampEn, which estimate the complexity at a single scale and may yield to contradictory and misleading findings. Multiscale entropy (MSE) was proposed by Costa et al. [4] to estimates the complexity of time series at multiple scales and had been one of the most effective approaches to reveal the inherent complexity of time series over multiple time scales. However, MSE may cause undefined entropy when the length of time series is too short. By using fuzzy entropy that widely used in various fields [7-12] replacing sample entropy [5, 6], and multi-scale fuzzy entropy (MFE) was proposed and successfully used to the fault diagnosis of rolling bearing [13].

However, with the increasing of scale factor, the length of coarse grained time series will decrease, which will result in the MFE curve fluctuating at the larger scale factors. To overcome the defect of MFE, the Sigmoid-based refined composite multi-scale fuzzy entropy (SRCMFE) is developed to measure the complexity of time series over multiple scales. By comparing SRCMFE with MFE, we find that SRCMFE is more reasonable and has stronger relative consistency and less dependency on the data length. Thus in this paper, SRCMFE is utilized to extract complexity features from vibration signals of rolling bearing over different temporal scales.

The fault features obtained by SRCMFE generally contains too many entropies in different scales. The features often have high dimension and contain redundant information, which cannot be able to reflect the fault features precisely and results in a lower fault identifying rate. A feature dimension reduction method is needed to refine the extracted features and improve the fault diagnosis efficiency of rolling bearing. Manifold learning is a kind of dimension reduction method that based on the concept of topological manifold and mainly includes linear and nonlinear manifold learning algorithms. t-distributed stochastic neighbor embedding (t-SNE) manifold learning algorithm recently introduced by Laurens et al [14] is a nonlinear visualizing dimensionality reduction algorithm with deep learning, by using which a low dimensional manifold structure can be recovered from high dimensional data. Hence, t-SNE is utilized to reduce dimension of the features. After that, a multi-fault classifier is needed for an automatic working condition identification of rolling bearing. The most often used methods in mechanical fault diagnosis fields mainly include artificial neural networks [15] and support vector machines [16]. However, they both ignore the intrinsic relationship among the extracted features. In fact, there are some inherent variable relations more or less in the features. The recently proposed classification method called variable predictive models based on class discrimination (VPMCD) is to fully make use of the intrinsic relationship between the features [17]. In VPMCD method, firstly the mathematical prediction models for revealing the intrinsic relationships of features are established for the different categories. Then the feature values of the tested samples are predicted by using these mathematical models and the squares sum of prediction error is taken as a criterion for judging classification and further pattern recognition. Therefore, VPMCD is employed to the rolling bearing fault diagnosis and a new fault diagnosis approach based on RCMFE, t-SNE and VPMCD is proposed for fault diagnosis of rolling bearing.

The rest of this paper is organized as follows. The proposed SRCMFE method is given in section 2. Comparison study of SRCMFE with MFE is given in Section 3. A novel rolling bearing fault diagnosis approach is put forward in Section 4. The conclusion is given in the final section.

## 2. Sigmoid-based refined composite multi-scale fuzzy entropy related methods

### 2.1. *Improved fuzzy entropy*

(1) For a given time series $\{u = u(1), u(2), u(3), ..., u(i), ...u(N), 1 \le i \le N\}$ with length *N*, the embedding dimension *m* and similar tolerance *r*. The vectors $X_i^m = \{u(i), u(i+1), ..., u(i+m-1)\} - u_0(i), i = 1, 2, ..., N-m+1\}$ can be formed, in which $X_i^m$ indicates that *m* consecutive *u* values, as the *i*th point and generalized by removing their baseline

$$u_0(i) = m^{-1}\sum_{k=0}^{m-1} u(i+k) \tag{1}$$

(2) The distance between $X_i^m$ and $X_j^m$ is defined as

$$d_{ij}^m = d[X_i^m, X_j^m] = \max_{k\in(0,m-1)}\{|[u(i+k)-u_0(i)]-[u(j+k)-u_0(j)]|\},\ i,j=1,2,...,N-m, i\neq j \tag{2}$$

(3) The similarity degree $D_{ij}^m$ of $X_i^m$ and $X_j^m$ is calculated by the fuzzy function $\mu(d_{ij}^m, a, r)$ as

$$D_{ij}^m = \mu(d_{ij}^m, a, r) = (1+e^{-a(d_{ij}^m - r)})^{-1} \tag{3}$$

(4) The function $\varphi^m$ is defined as

$$\varphi^m(n,r) = \frac{1}{N-m}\sum_{i=1}^{N-m}\left(\frac{1}{N-m-1}\sum_{\substack{j=1\\ j\neq i}}^{N-m} D_{ij}^m\right) \tag{4}$$

Similarly, $\varphi^{m+1}(n,r)$ is also achieved as

$$\varphi^{m+1}(n,r) = \frac{1}{N-m}\sum_{i=1}^{N-m}\left(\frac{1}{N-m-1}\sum_{\substack{j=1\\ j\neq i}}^{N-m} D_{ij}^{m+1}\right) \tag{5}$$

(5) The FuzzyEn of time series is defined as the negative natural logarithm of the deviation of $\varphi^m$ from $\varphi^{m+1}$, $FuzzyEn(m,n,r) = \lim_{N\to\infty}[\ln\varphi^m(n,r) - \ln\varphi^{m+1}(n,r)]$, and for finite datasets, it can be estimated by the statistic,

$$FuzzyEn(m,r,N) = \ln\varphi^m(r) - \ln\varphi^{m+1}(r) \tag{6}$$

In the original FuzzyEn [5] the exponential function $\mu(d_{ij}^m, n, r) = e^{-(d_{ij}^m)^n/r}$ was used and in literature [13] $\mu(d_{ij}^m, n, r) = e^{-\ln 2(d_{ij}^m/r)^n}$ used to measure the similarity of two vectors. In this paper the exponential functions are replaced by Sigmoid-based membership function $\mu(d_{ij}^m, a, r) = (1+e^{-a(d_{ij}^m - r)})^{-1}$.

The Sigmoid-based membership function is a Sigmoid-based function of $x$, which depends on parameters $a$ and $r$ is given by:

$$f(x;a,r) = (1+e^{-a(x-r)})^{-1} \tag{7}$$

Figure1 illustrates the Sigmoid-based membership function with $r$=0.5 for different values of $a$. From Figure 1, we can clearly observe that the fuzzy membership function changes gradually and continuously, which is unlike the Heaviside function utilized in SampEn that changes abruptly.

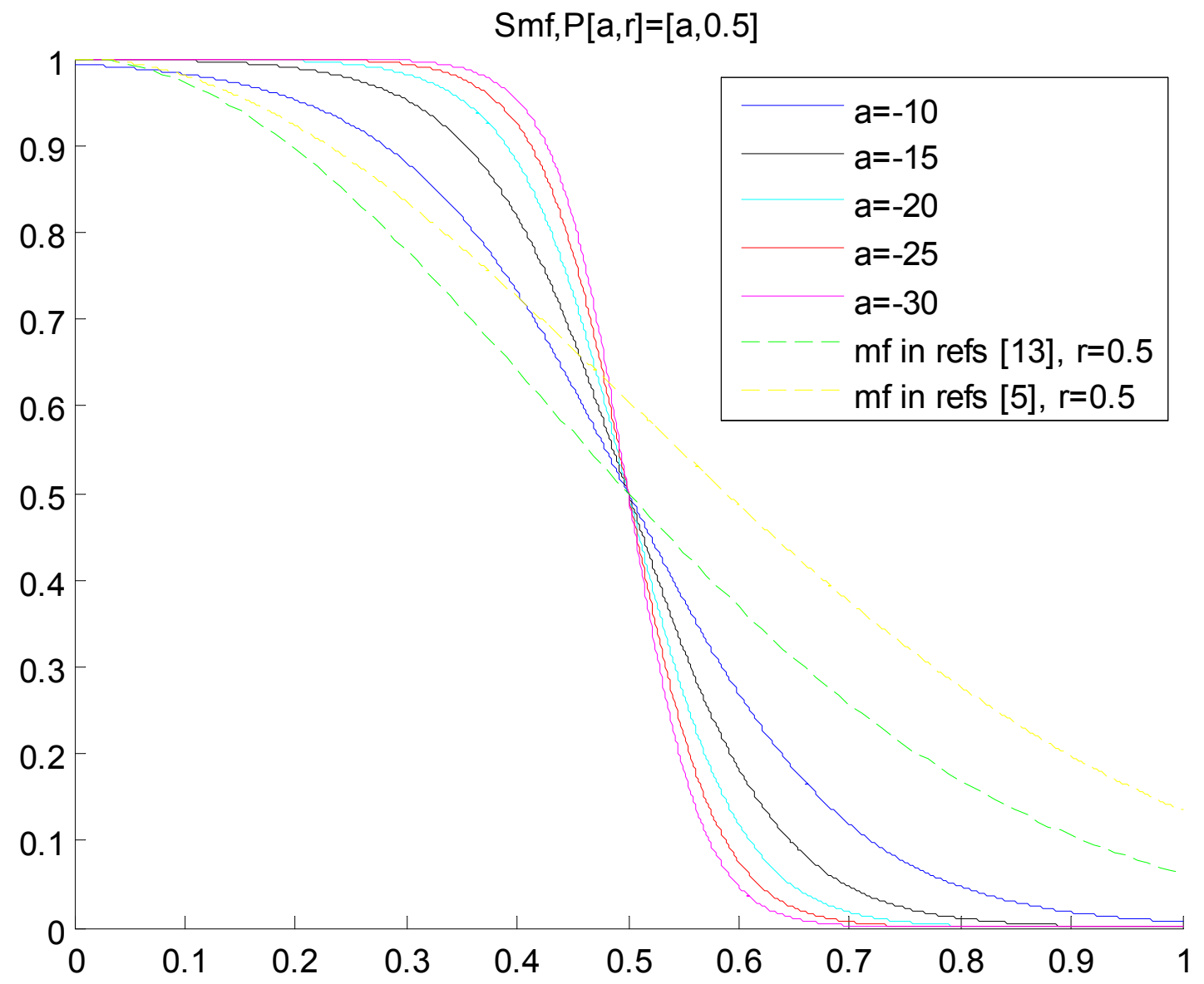

Figure1. The influence of parameter *a* on the shape of membership function

2.2. *Sigmoid-based multiscale fuzzy entropy*

The algorithm of Sigmoid-based multi-scale fuzzy entropy (SMFE) mainly contains two steps.

(1) Given a time series $\{x(i), 1 \le i \le N\}$, coarse grained time series $\{y_\tau\}$ is established as

$$y_\tau(j) = \frac{1}{\tau} \sum_{i=(j-1)\tau+1}^{j\tau} x(i), 1 \le j \le N / \tau \tag{8}$$

Where $\tau$ is the scale factor, to achieve $\tau$ coarse-grained time series over coarse-grained process, the original time series is divided into non-overlapping windows with length $\tau$ and the data points segmented in each window are average.

(2) FuzzyEn is calculated for each coarse-grained time series and then the SMFE at scale factor $\tau$ is defined as

$$SMFE(y, m, r, \tau) = FuzzyEn(y_\tau, m, r) \tag{9}$$

Since the length of each coarse-grained time series is equal to the original time series divided by the scale factor $\tau$, the variance of FuzzyEn will increases with the decrease of length of the coarse-grained time series. To overcome this, the Sigmoid-based composite multiscale fuzzy entropy (SCMFE) algorithm is proposed in the subsection.

2.3 *Sigmoid-based composite multiscale fuzzy entropy*

The algorithm of SCMFE can be given as follows.

(1) For scale factor $\tau$, the original time series $\{x(i), 1 \le i \le N\}$ can be divided into $\tau$ coarse-grained signals. The $\varepsilon$-th coarse-grained time series $y_{\varepsilon}^{(\tau)} = \{y_{\varepsilon,1}^{(\tau)}, y_{\varepsilon,2}^{(\tau)}, ..., y_{\varepsilon,\zeta}^{(\tau)}\}$ of original time series is calculated as Formula (10) [18, 19]

$$y_{\varepsilon,j}^{(\tau)} = \frac{1}{\tau} \sum_{i=(j-1)\tau+\varepsilon}^{j\tau+\varepsilon-1} x_i, 1 \le j \le N/\tau, 1 \le \varepsilon \le \tau \tag{10}$$

(2) The FuzzyEn values of all coarse-grained time series are calculated by the SCMFE algorithm over scale factor $\tau$, and the SCMFE is achieved as the means of $\tau$ FuzzyEns. Hence, for scale factor $\tau$, the SCMFE is defined as Formula (11)

$$SCMFE(x,m,r,\tau) = \frac{1}{\tau}\sum_{\varepsilon=1}^{\tau} FuzzyEn(y_{\varepsilon}^{(\tau)}, m, r) = \frac{1}{\tau}\sum_{\varepsilon=1}^{\tau}\left(-\ln \frac{n_{\varepsilon,\tau}^{m+1}}{n_{\varepsilon,\tau}^{m}}\right) \tag{11}$$

where $m_{\varepsilon,\tau}^{m}$ represents the total amount of the $m$-dimensional matched vector pairs.

In SCMFE algorithm, if one of the values $n_{\varepsilon,\tau}^{m+1}$ or $n_{\varepsilon,\tau}^{m}$ is zero, the SCMFE value is undefined. Therefore, when SCMFE is utilized to analyze shorter time series, it may cause undefined entropy. Hence, once the SCMFE algorithm is employed to analyze short time series, some entropy will have no definitions. To overcome the shortcomings of SCMFE, the SRCMFE is proposed.

2.4 *Sigmoid-based refined composite multiscale fuzzy entropy*

The SRCMFE can be given as follows.

(1) The formula described in (10) is employed to obtain the coarse-grained signals over various scale factors. For scale factor $\tau$, $m_{\varepsilon,\tau}^{m+1}$ and $m_{\varepsilon,\tau}^{m}$, the amount of matched vector pairs, are achieved for different $\tau$ coarse-grained time series [20].

(2)For a given embedding dimension $m$ and scale factor $\tau$, $m_{\varepsilon,\tau}^{m+1}$ & $m_{\varepsilon,\tau}^{m}$ for each $x_{\varepsilon}^{(\tau)}$ are obtained. The SRCMFE approach is showed as Formula (12)

$$SRCMFE(x,m,r,\tau) = -\ln\left(\frac{\overline{n}_{\varepsilon,\tau}^{m+1}}{\overline{n}_{\varepsilon,\tau}^{m}}\right) \tag{12}$$

where $\overline{n}_{\varepsilon,\tau}^{m+1} = \frac{1}{\tau}\sum_{\varepsilon=1}^{\tau} n_{\varepsilon,\tau}^{m+1}$ and $\overline{n}_{\varepsilon,\tau}^{m} = \frac{1}{\tau}\sum_{\varepsilon=1}^{\tau} n_{\varepsilon,\tau}^{m}$.

SRCMFE has overcome the drawback of SCMFE and only is undefined when $\overline{n}_{\varepsilon,\tau}^{m+1}$ or $\overline{n}_{\varepsilon,\tau}^{m}$ is equal to zero. Therefore, compared with the SCMFE, SRCMFE has stronger capability in cutting down the probability of inducing undefined entropy.

## 3. Comparison study

To illustrate the effectiveness of proposed method, SRCMFE and MFE are used to analyzing two random noise signals, white noises and 1/f noises in this section.

In order to study the statistical behaviors of the MFE and SRCMFE methods further, white and 1/f noises, which as two widely used signals tested, are analyzed here by using MFE and SRCMFE. There was 200 independent noise samples used in each length of the data. Fig.2 shows the time domain waveform of white noise and 1/f noise. Fig.3 (a) shows the FuzzyEn curve of white noise with different length over multiple scales, which decreases monotonically with larger fluctuation than the curve of SRCMFF showed in Fig.3 (b). As for 1/f noise, Fig.3 (c) shows how the fluctuation of the FuzzyEn curve obtained with different length of the 1/f noise by using MFE algorithm is remarkably larger than that obtained by using SRCMFE algorithm showed in Fig.3(d). Besides, the entropy values calculated by using SRCMFE remain better stability and consistency than the MFE over multiple scales.

To verify the accuracy of the MFE and SRCMFE methods, MFE and SRCMFE of white noise and 1/f noise were calculated with 1000 data points and 10000 data points in each noise sample with 200 analysis. Fig.4 present the standard deviations (SDs) of entropies of white noise and 1/f noise with 200 analysis samples obtained by using MFE and SRCMFE algorithm. Regarding white noise, Fig.4 (a) the SD of the MFE was greater than that of the SRCMFE method. In the case of 1/f noise, Fig.4 (b) shows that the SDs of the entropy value obtained by using MFE was larger than those obtained by using SRCMFE method.

When the data length is increased to 10000, results obtained of SDs from white noise and 1/f noise by using MFE and SRCMFE showed in Fig.4 (c) and Fig.4 (d) are the same as in Fig.3 (a) and (b), apart from the latter's the standard deviation of entropy is much smaller. Hence, In comparison with the estimation by the MFE, the SDs of the entropy estimator can be improved by the SRCMFE evidently.

The above analysis results indicate that the validities of the MFE decreased as the fluctuation of the entropy of the time series increased over multiple scales. Therefore, the shorter the time series, the lower the validities of the MFE algorithm is. In addition, SRCMFE is independent on data length, even if as the decrease of the length of time series, to a large extent, the SD values estimated by SRCMFE are is much smaller than those estimated by MFE. Therefore, the entropy estimated by the SRCMFE algorithm is more precise and consistent than those estimated by the MFE algorithm. Hence, the SRCMFE algorithm is superior to MFE algorithms.

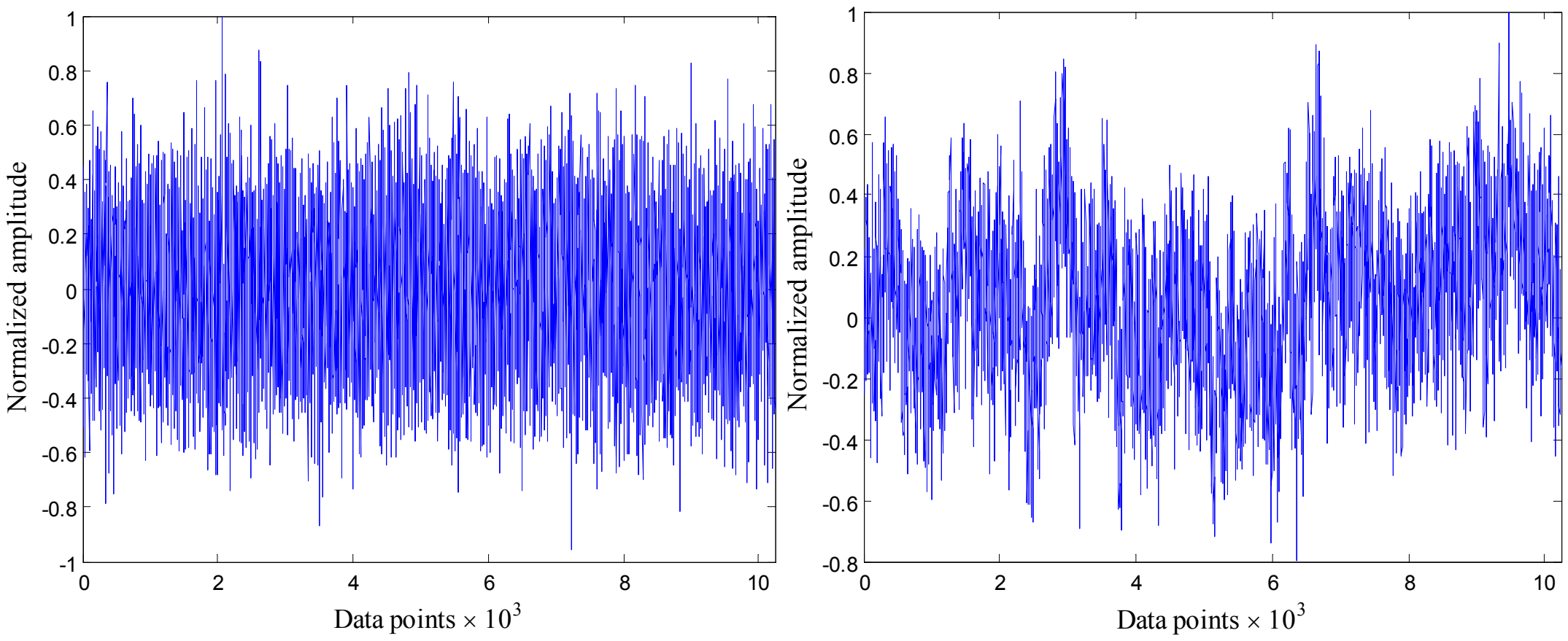

(a) Waveforms of Gaussian white noise (b) Waveforms of Gaussian 1/f noise

Fig.2 Waveforms of Gaussian white noise and 1/f noise

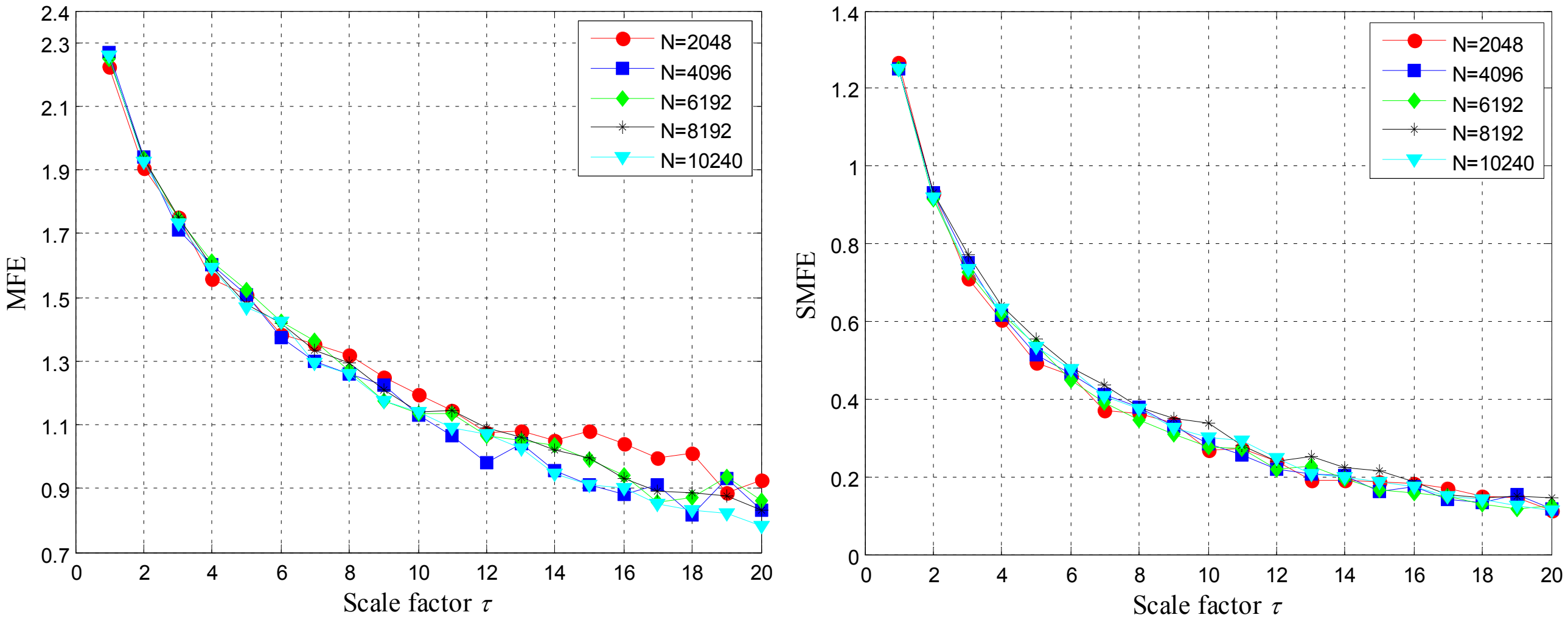

(a) MFE of white noise with different length (b) SMFE of white noise with different length

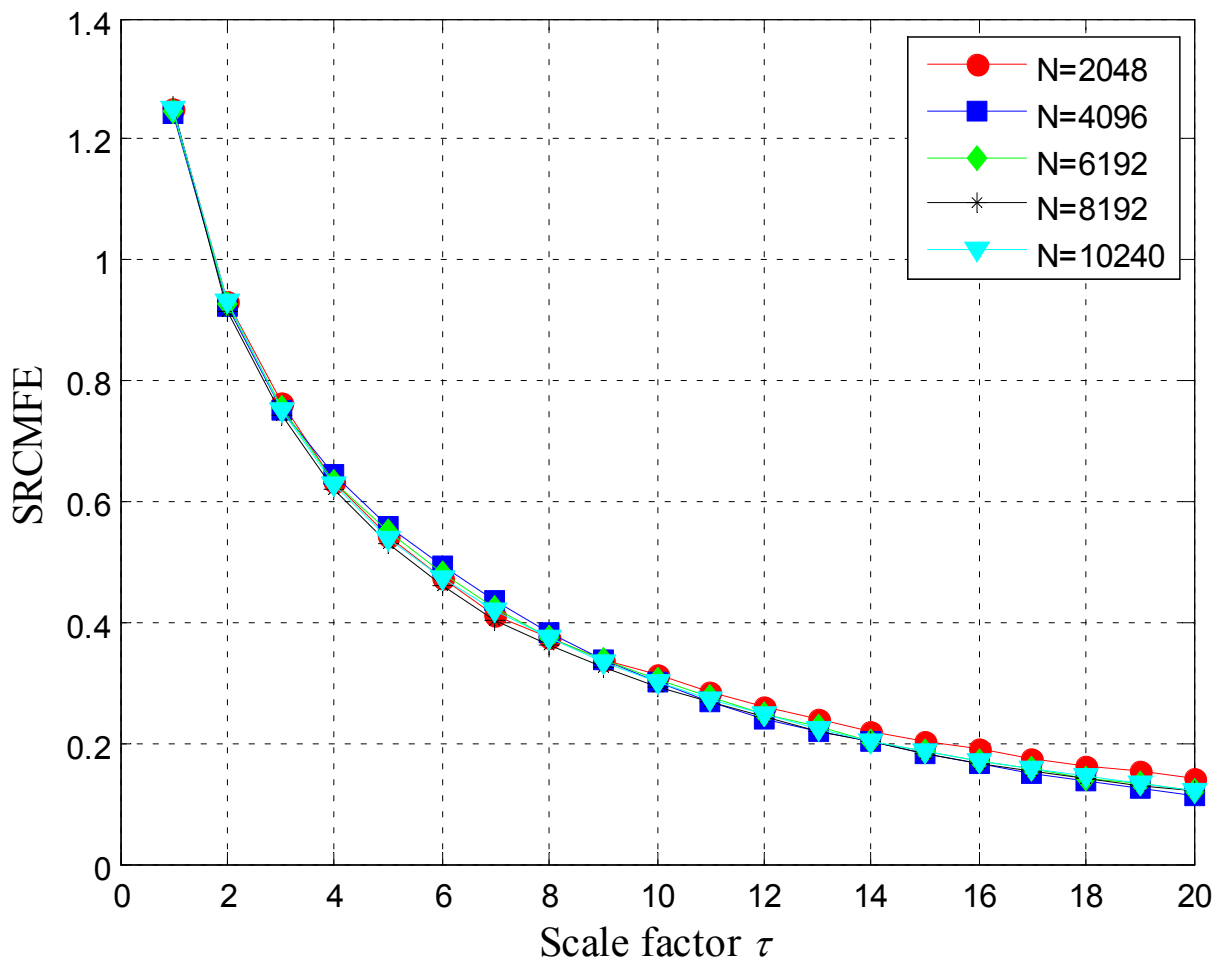

(c) SRCMFE of white noise with different length

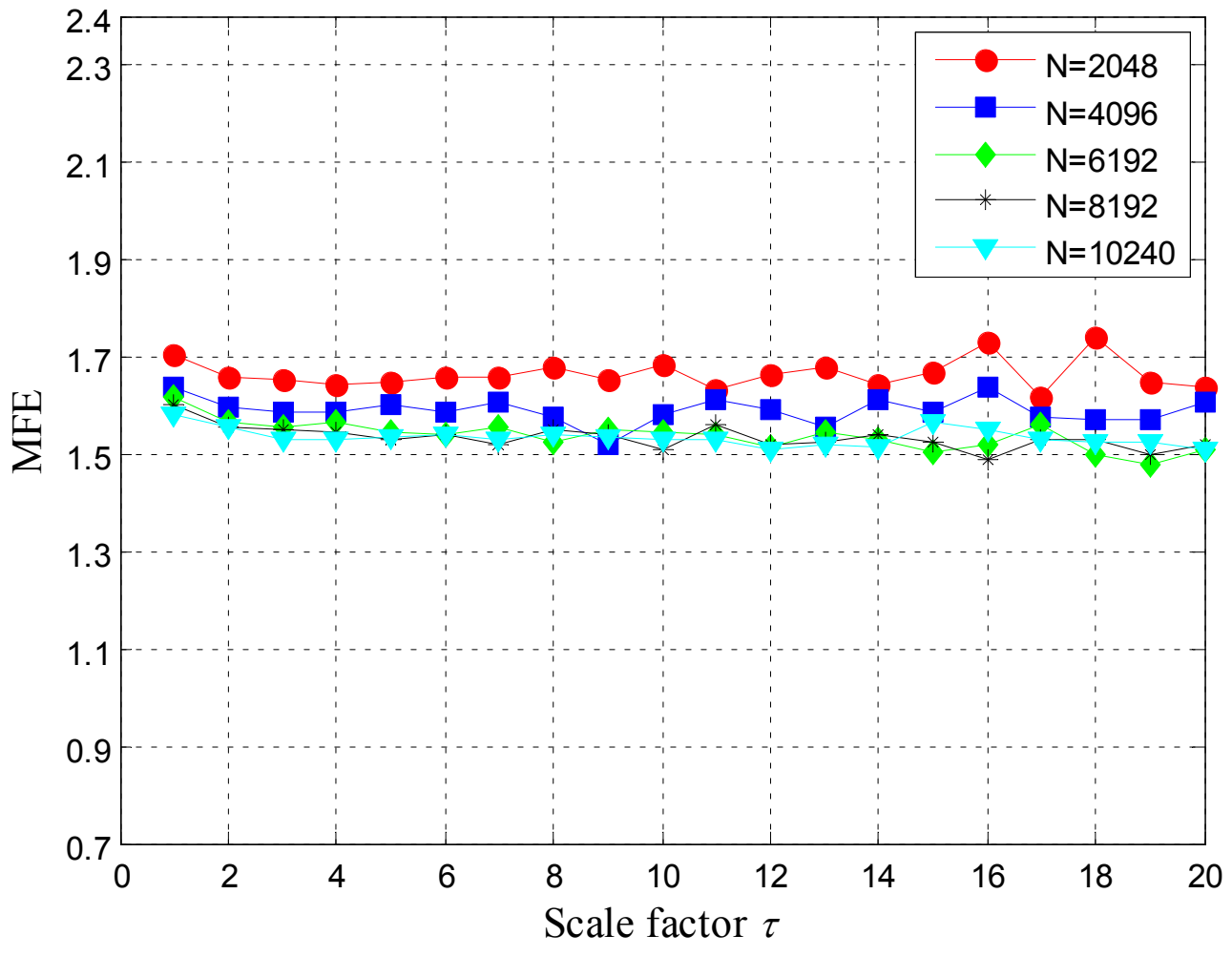

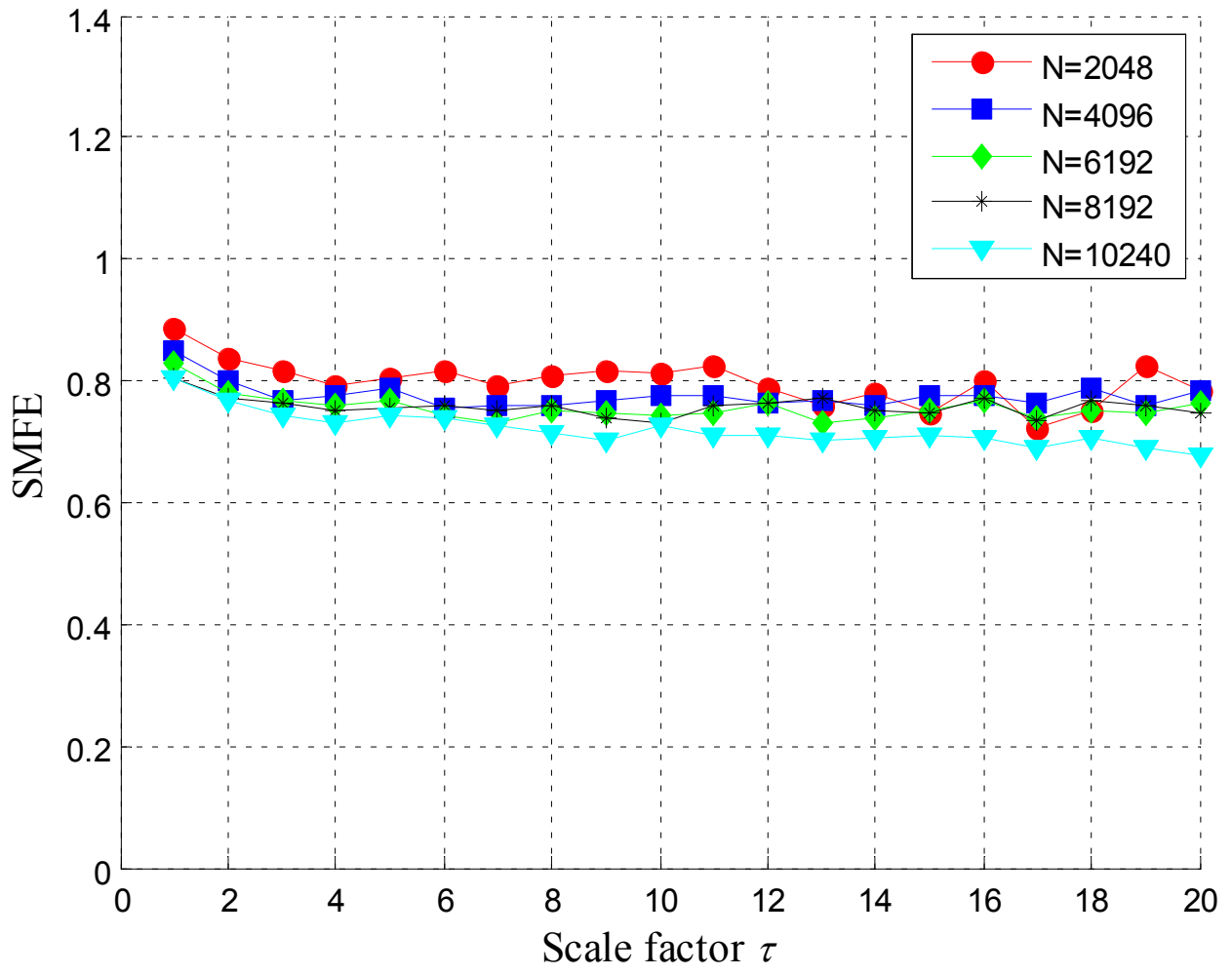

(d) MFE of 1/f noise with different length

(e) SMFE of 1/f noise with different length

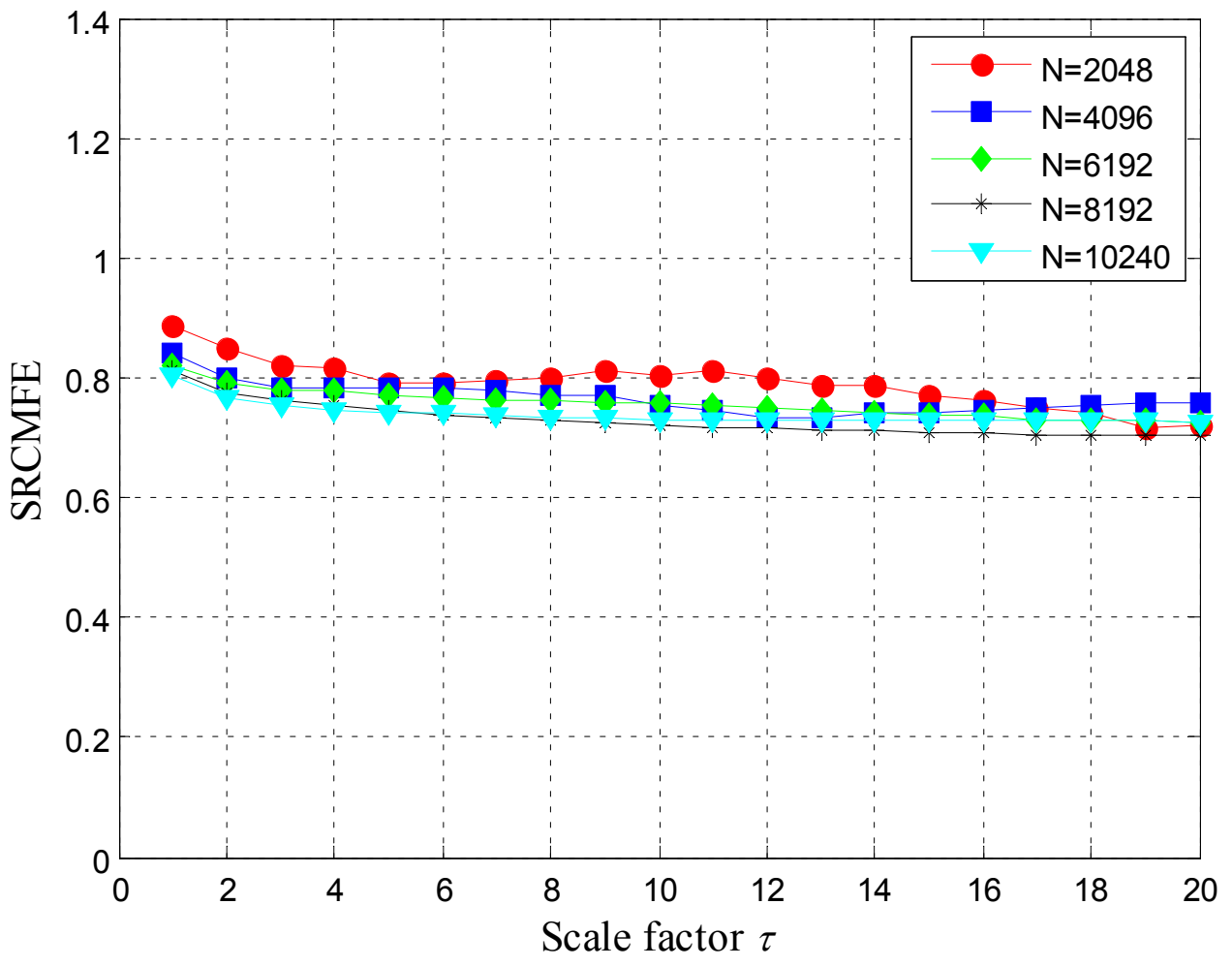

(f) SRCMFE of 1/f noise with different length

Fig.3 MFE, SMFE and SRCMFE of Gaussian white noise and 1/f noise with

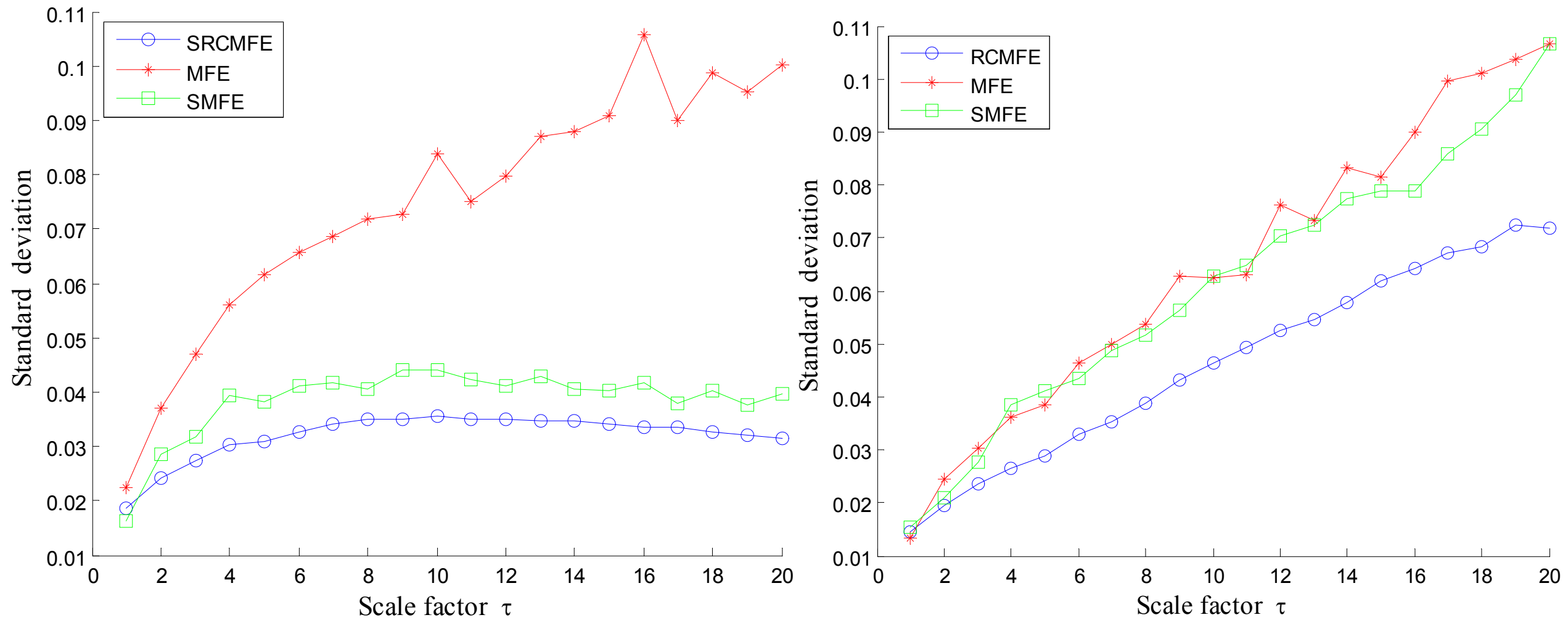

(a) White noise with 1000 data points (b) 1/f noise with 1000 data points

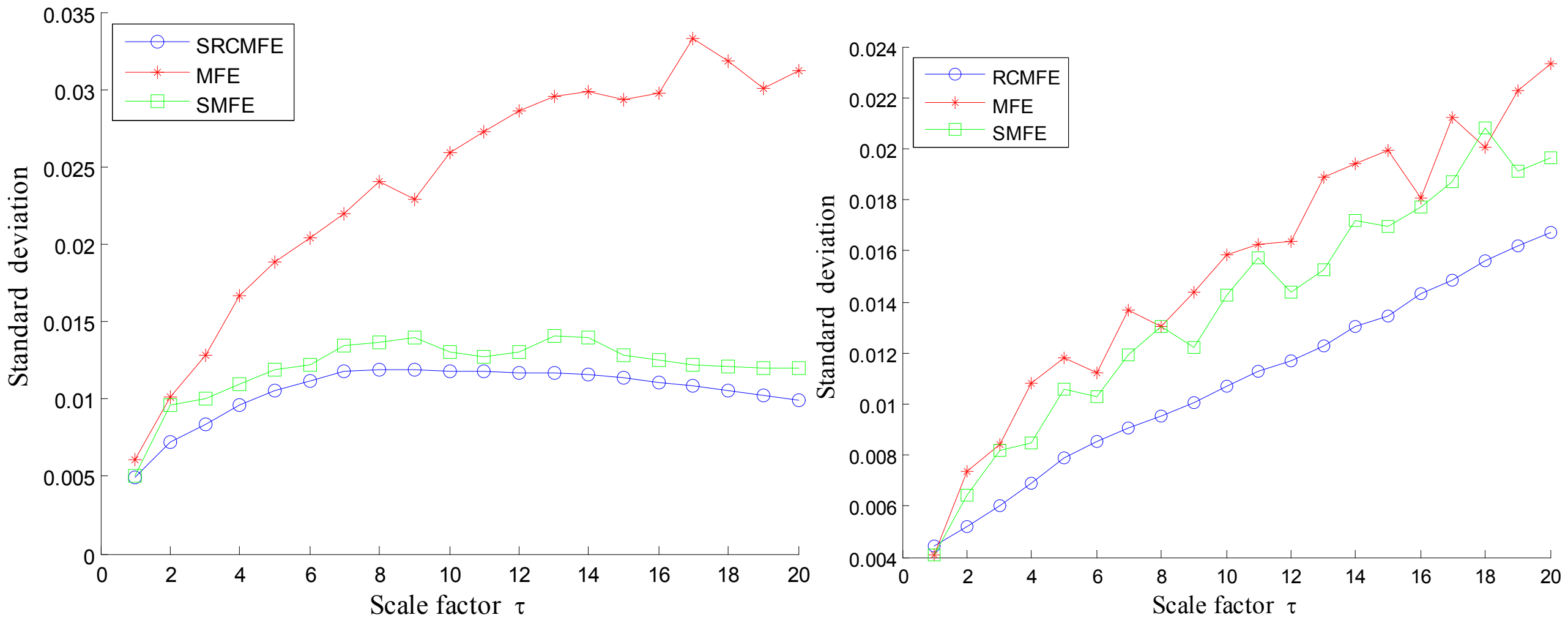

(c) White noise with 10000 data points (d) 1/f noise with 10000 data points

Fig.4 Standard deviation of 200 analysis results of white noise and 1/f noise.

## 4. The proposed fault diagnosis approach and its application

### *4.1 t-SNE algorithm*

When we have extracted MDEs from vibration signals of rolling bearing, it is necessary to select the most important sensitive features to construct the sensitive fault features for intelligent fault diagnosis. In this paper t-SNE is utilized to reduce feature dimensions and its mainly steps are described as follows.

(1) For an original data sequence $X=\{x_1,x_2,...,x_n\}$, the joint probabilities $p_{ij}$ are defined to measure the pairwise similarity between objects $x_i$ and $x_j$. The pairwise affinities $P_j\,|\,i$ with perplexity (*Perp*) are calculated according to Eq. (13). The perplexity *Perp* is as a cost function parameter.

$$P_j \mid i = \frac{\exp\left(-\frac{\| x_i - x_j \|^2}{2\sigma_i^2}\right)}{\sum_{k \neq i} \exp\left(-\frac{\| x_i - x_k \|^2}{2\sigma_i^2}\right)} \tag{13}$$

Where $\sigma_i$ is Gauss variance of data point $x_i$.

(2) Once the data point $x_i$ is an outlier, it will cause that the position of map point is not well determined by the positions of the other map points. To deal with this issue, the joint probabilities $p_{ij}$ are defined in the high-dimensional space as symmetry conditional probabilities. So we set $p_{ij} = \frac{p_j \mid i + p_i \mid j}{2n}$, where *n* is the total number of data points.

(3) Let the mapping points of the high-dimensional space data points $x_i$ and $x_j$ in the low-dimensional space are $y_i$ and $y_j$. In order to satisfy $p_{ij} = q_{ij}$, where $q_{ij}$ is the joint probabilities in low dimensional space, then the distance in the low-dimensional space should be slightly smaller for the points closer in the high-dimensional space. Also the distance in the low-dimensional space should be farther for the points that are far apart in the high-dimensional space. Hence, the joint probabilities $q_{ij}$ in low dimensional space are defined as Eq. (14) by using a Student t-distribution with one degree of freedom in t-SNE.

$$q_{ij} = \frac{\left(1 + \| y_i - y_j \|^2\right)^{-1}}{\sum_{k \neq l} \left(1 + \| y_k - y_l \|^2\right)^{-1}} \tag{14}$$

(4) To measure the similarity between high-dimensional space joint conditional probability distribution P and low-dimensional space joint conditional probability distribution Q, and by gradient descent algorithm minimizing cost function $C = \sum_i KL(P_i \| Q_i) = \sum_i \sum_j p_{j|i} \log \frac{p_{j|i}}{q_{j|i}}$ that Kullback-Leibler divergence between *P* and *Q*, the gradient $\delta C / \delta y_i$ is calculated according to Eq. (15)

$$\frac{\delta C}{\delta y_i} = 4 \sum_j \left(p_{ij} - q_{ij}\right)\left(y_i - y_j\right)\left(1 + \| y_i - y_j \|^2\right)^{-1} \tag{15}$$

(5) Low dimensional data can be obtained according to Eq. (16)

$$y^{(t)} = y^{(t-1)} + \eta \frac{\delta C}{\delta y} + \alpha(t)\left(y^{(t-1)} - y^{(t-2)}\right) \tag{16}$$

where learning rate $\eta$ and momentum $\alpha(t)$ are optimization parameters.

(6) Iterate loop stepts (3) to (5) until $t$ from 1 to $T$, where $T$ is maximum number of iterations that should be pre-set. Finally low dimensional data $y^{(T)} = \{y_1, y_2, ..., y_n\}$ are obtained.

The cost function in t-SNE algorithm is different from SNE in two aspects: (1) t-SNE using symmetric SNE cost function of reduced gradient; (2) It uses the t- student distribution instead of the Gaussian distribution to calculate the similarity between two points in a low dimensional space. The heavy tailed distribution in low dimensional space is employed in t-SNE to slow down the aggregation and optimization of SNE. t-SNE as a nonlinear dimensionality reduction algorithm for deep learning, the structure of low dimensional manifold can be recovered from high-dimensional data, hence it can achieve dimensionality reduction and data visualization.

*4.2. The proposed fault diagnosis method*

Based on the superiority of SRCMFE, t-SNE and VPMCD, a novel bearing fault diagnosis method is proposed and its steps can be summarized as follows.

(1) Assume that the running states of rolling bearing contain K states and each state has N samples.

(2) SRCMFE of all vibration signal are computed and the maximum scale factor $\tau_{\max}$ feature values were obtained to represent the fault information of the vibration signals of rolling bearing in each group, the feature vector matrix is constituted: $R^{N\times\tau_{\max}}$, where $\tau_{\max}$ is the largest scale factor.

(3) t-SNE manifold learning algorithm is used to reduce the dimension of feature vector matrix and a low dimensional sensitive feature set $R^{N\times i}$ is obtained.

(4) $N$/2 groups training samples of each state are input to the VPMCD based multi-classifier for training and building the predictive model $VPM_i^k$, where $k$=1,2,... , $g$ represents different categories, $i$=1,2,..., $p$ represents different characteristic values.

(5) The rest samples are seen as test ones to test the trained VPMCD multi-classifier. The outputs of the VPMCD classifier are used to determine the working condition and fault type of the rolling bearing for an automatic fault diagnosis.

The flowchart of the proposed method can be summarized in Fig.5.

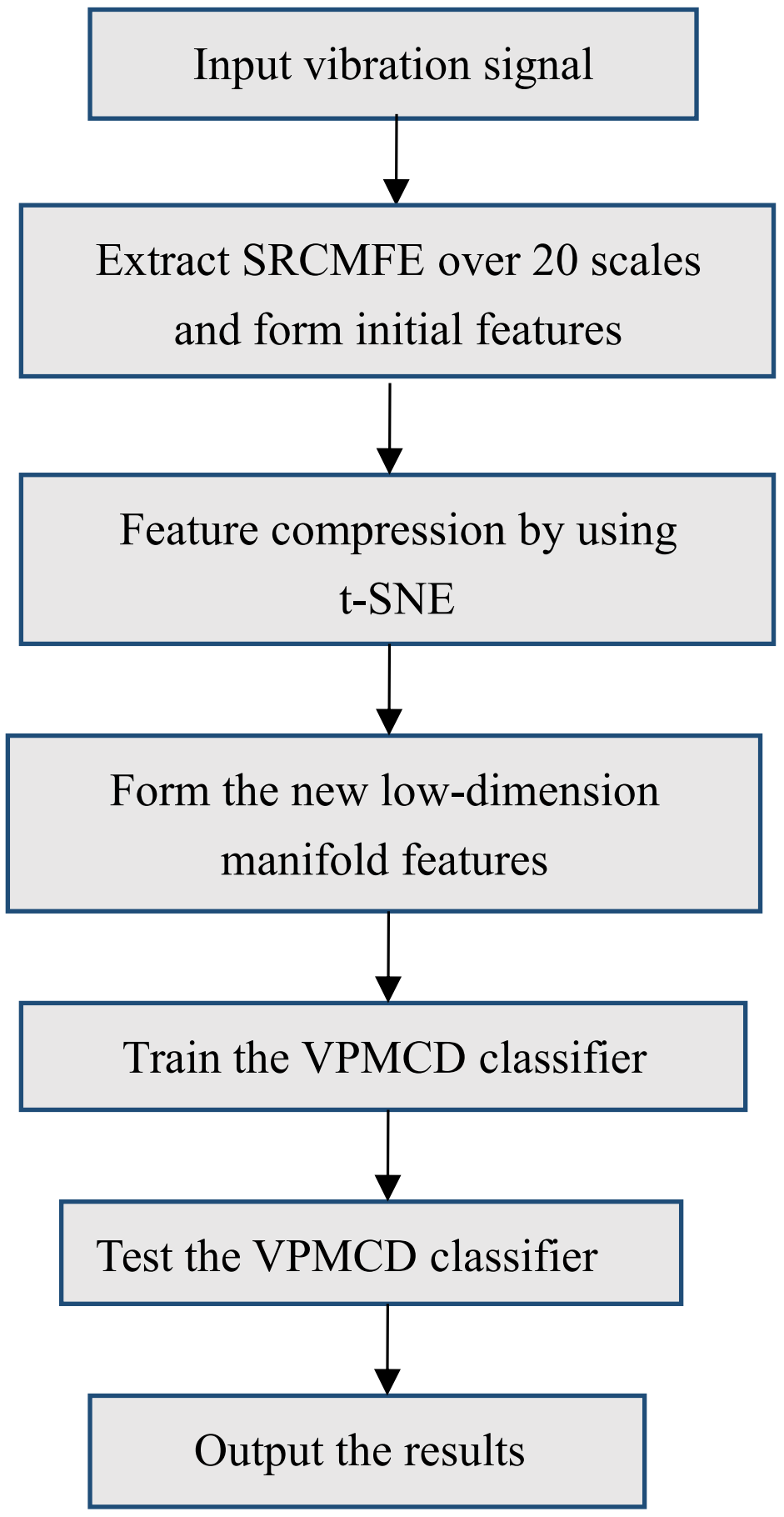

Fig.5 Flow chart of the proposed method

### *4.2. Experiment data analysis*

In order to validate the effectiveness of the SRCMFE algorithm and the proposed method for rolling bearing fault diagnosis. The proposed approach is applied to analyze the experimental data, which are kindly provided by Case Western Reserve University (CWRU) Bearing Data [21]. The experiment system are given in Fig.6, in which the 6205-2RS JEM SKF deep groove ball bearing was used in the test, the test bearings with electro-discharge machining fault are used to detect single point faults. The test stand is composed of a 2-horsepower motor and a dynamometer, which are connected by a torque transducer. An accelerometer with a bandwidth up to 5000 Hz was mounted on the motor housing at the drive end of the motor to collect the vibration signals from the bearing. The data collection system consisted of a high-bandwidth amplifier particularly designed for vibration signals and a data recorder with a sampling frequency of 12,000 Hz per channel, the motor revolving speed was 1, 730 r/min. The vibration signals of bearing were collected under seven kinds of conditions including the normal condition; inner race fault conditions (IRF), which are 0.1778 and 0.5334 mm in diameter; the outer race fault condition (ORF), which are 0.1778 and 0.5334 mm in diameter located at 6 o'clock positions and the ball fault condition (BF), which are 0.1778 and 0.5334 mm in diameter. In consideration of multi-fault categories and severities, the experiment turns into a seven-class classification issue. The data set totally consist of 203 data samples, and length of the each data sample is 4,096. Time waveform and their frequency spectrum of vibration signals seven fault categories are depicted in Fig.7 (a) and Fig.7 (b), respectively. The detailed description of the data set used in this paper is shown in Table 1.

It can be seen that it is difficult to discriminate among different rolling bearing conditions from time domain and their frequency spectrum. In addition, it is also untrustworthy to make classification results on the basis of the time domain waveforms and frequency spectrum. In general, the vibration signals of mechanical system with fault are nonlinear and non-stationary. Sigmoid-based RCMFE, as a nonlinear dynamic parameter estimation method, is utilized to extract the hidden fault characteristic information over multiple scales.

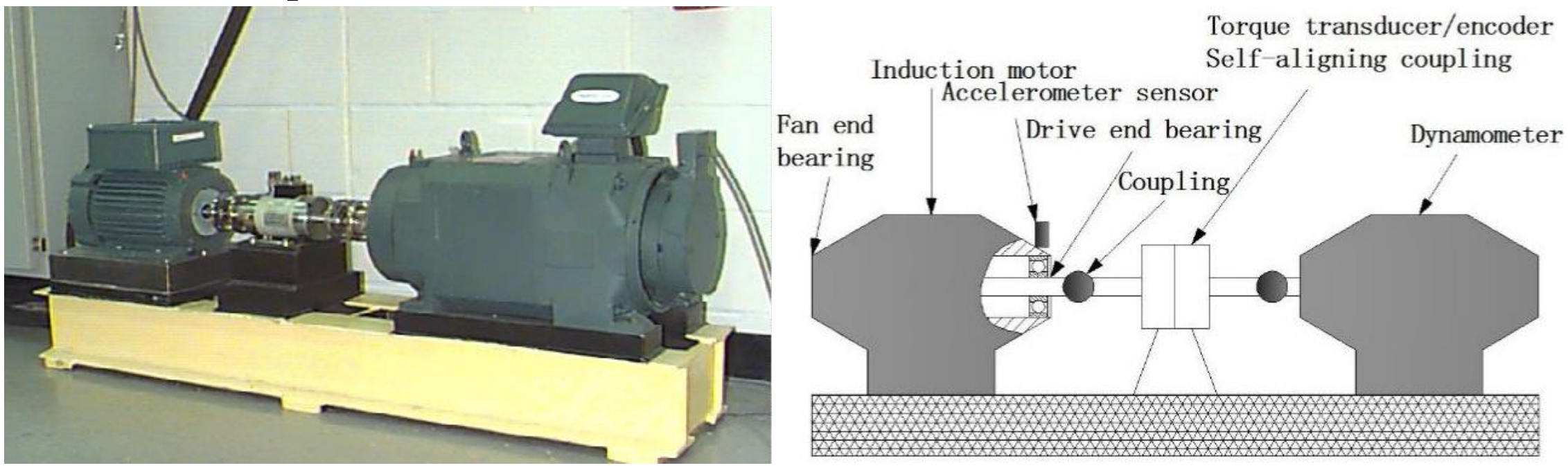

Fig.6 The rolling bearing experiment system and its sketch

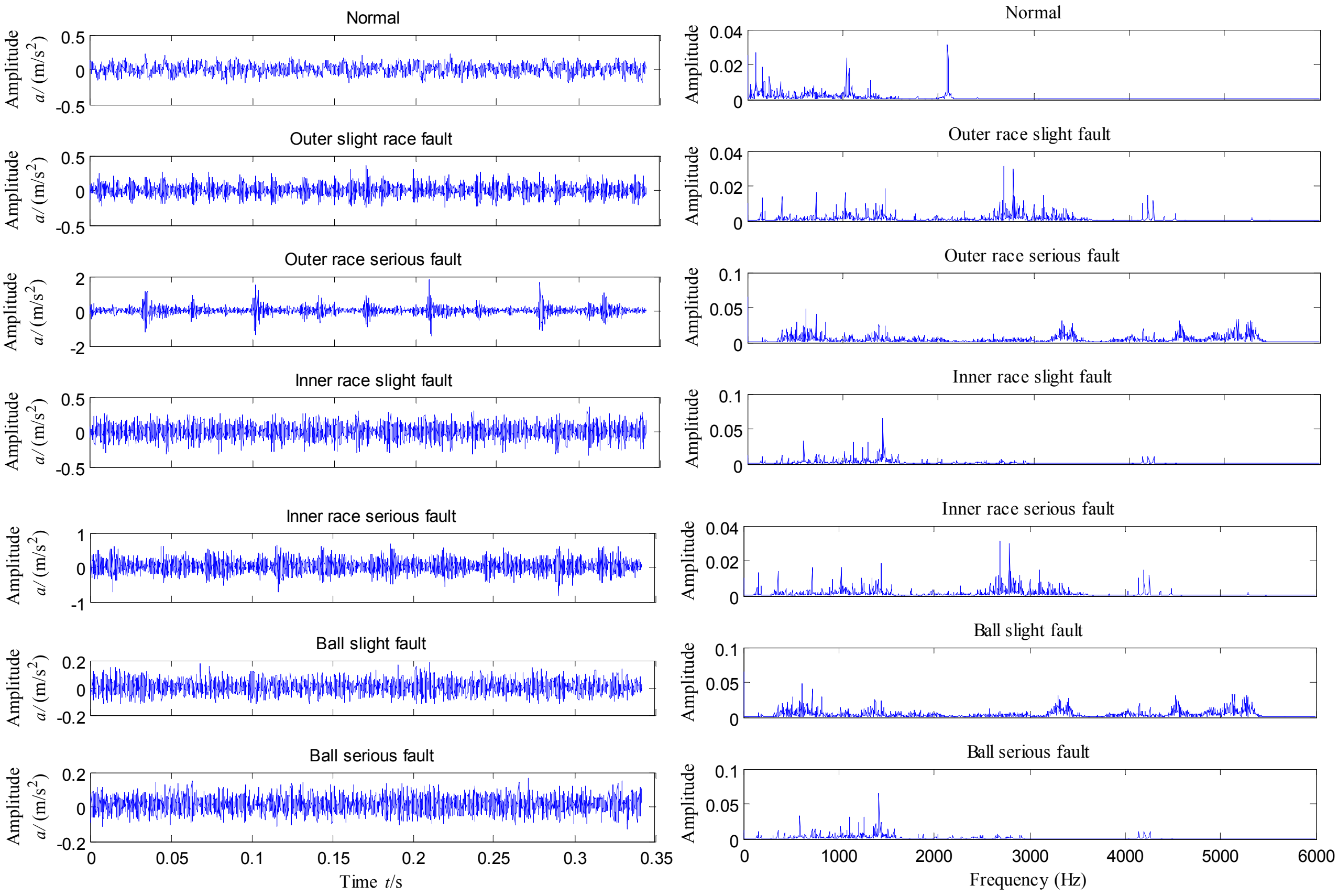

(a) Time waveform of vibration signals (b) Frequency spectrum of vibration signals

Fig.7 Time waveform (a) and their frequency spectrum (b) of vibration signals under seven fault categories

Table 1 Description of experimental data

| Fault class | Fault size (mm) | Fault severity | Number of training data | Number of testing data | Class label |
|---|---|---|---|---|---|
| Normal | | | 15 | 14 | 1 |
| ORF1 | 0.1778 | Slight | 15 | 14 | 2 |
| ORF2 | 0.5334 | Severe | 15 | 14 | 3 |
| IRF1 | 0.1778 | Slight | 15 | 14 | 4 |
| IRF2 | 0.5334 | Severe | 15 | 14 | 5 |
| REF1 | 0.1778 | Slight | 15 | 14 | 6 |
| REF2 | 0.5334 | Severe | 15 | 14 | 7 |

The SRCMFE over 20 scales of rolling bearing vibration signals shown in Fig.8 (a) is given in Fig.7, from which can be seen that the SRCMFE of normal rolling bearing vibration signal has the largest entropy values over most scales. Hence, this result indicates that the vibration signals of normal status rolling bearing are more complex than those vibration signals of faulty rolling bearing. However, FuzzyEn of normal rolling bearing vibration signal is smaller than those of vibration signals with fault bearing in the first scale from Fig.8. This illustrates the long-term structures in the time series cannot be captured only by FuzzyEn.

Above analysis results able to explained by the fact that as the bearing works in a healthy condition, the vibration signal over most scales, in general more than 2, has stronger time irreversibility [22] and has lower self-similarity, and higher probability to generate new modes. However, above characteristic property will be the opposite change when the rolling bearing get out of order. From another aspect, the vibration of bearing will contains more impacting frequency components (frequency spectrum shown in Fig.7 (b)) once the rolling bearing failure. Therefore, it will leading to lower FuzzyEn values. This explained that the curve of SRCMFE for healthy rolling bearing lies above that of the fault rolling bearing.

Besides, once rolling bearing fault occurs, the vibration signal of mechanical system has obvious impact characteristics. Due to different location of the fault bring about the frequencies of different impacted, the complexity of the signal will be different. As the outer ring is fixed, the fault feature frequency is the main impact frequency. And the inner ring along with the shaft rotation, not only the inner ring fault characteristic frequency is the main impact frequency, but also in the vibration signal of inner ring fault will appear regular periodic shaft frequency, this will lead to an increase in the sparsity of the signal, hence the curve of defective inner ring bearing below that of defective outer ring bearing in most coarse grain scales. And the rolling element is not only autorotation but also around the axis movement, although the fault feature frequency components account for the main impact frequency, however, due to the non-fixed motion there will be a number of different impact frequency superposition leading to non-resonance impact, this will decrease the sparsity of the vibration signal, so the entropy values of rolling element vibration signals are higher than that of inner ring rolling bearing and outer ring bearing over multiple scales. Therefore, the curve of SRCMFE for defective rolling element vibration signals is above that of defective inner ring bearing vibration signals and defective outer ring bearing vibration signals. In addition, the greater the fault degree, the stronger the self-similarity and the sparsity of the vibration signal, the smaller the FuzzyEn entropy. This explains why the curve of SRCMFE for slight fault rolling bearing lies above that of the serious fault rolling bearing.

Although, SRCMFE able to be employed to measure the complexity of bearing vibration signals,

it remains very difficult to discriminate the 7 classes of identification issue from the curve of their corresponding SRCMFE. To achieve the fault diagnosis automation and improve the efficiency of decision-making, a novel VPMCD-based multi-classifier will be constructed and utilized to fulfill automated fault diagnosis.

However, high dimensional feature vectors inevitably there will be redundant and unrelated feature information with fault, and it will make VPMCD classifier time-consuming in the training. Therefore, it is so necessary to reduce the dimension of the high dimension feature vectors that improve the efficiency of VPMCD classifier. t-SNE, a nonlinear manifold learning algorithm of deep learning, here is utilized to achieve dimension reduction. It is an impactful dimension reduction tool. Fig.9 shows that two-dimensional manifold diagrams (Fig.9 (a)) and three-dimensional manifold diagrams (Fig.9 (b)) are obtained by using t-SNE algorithm after extracting SRCMFE of bearing vibration signals.

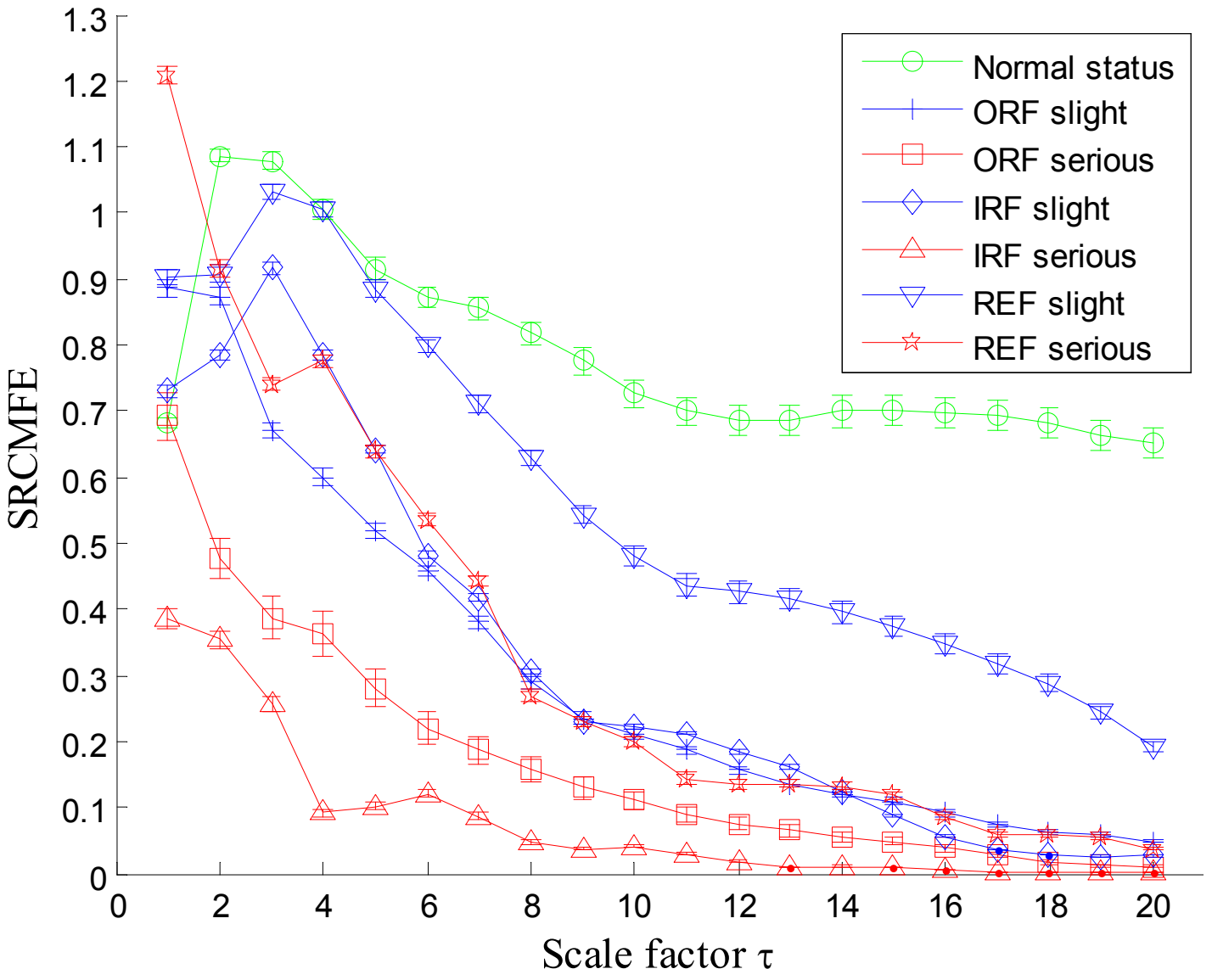

Fig.8 SRCMFE over 20 scales of the corresponding vibration signals

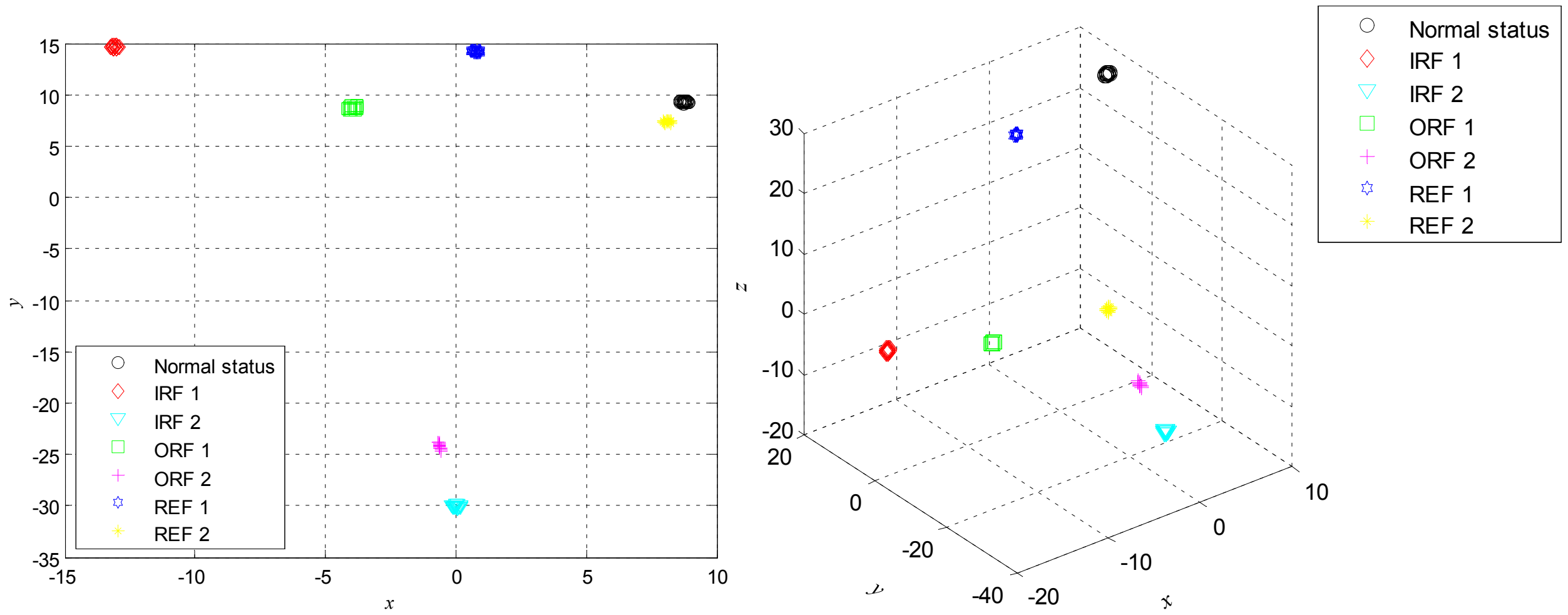

(a) Two-dimensional histogram (b) Three-dimensional histogram

Fig.9 Two-dimensional histogram (a) and three-dimensional histogram (b) by using t-SNE algorithm

It can be seen that different categories able to separate obviously from both the two-dimensional histogram and three-dimensional histogram. The new feature vectors are divided into a training data set (105 samples) and a testing data set (98 samples). And the set of training data is used to train the VPMCD classifier. After that, the VPMs are obtained and the testing data set is utilized to validate classification accuracy of VPMCD clasifier.

The classification results of the proposed approach are shown in Fig.10, which include the VPMCD outputs and the desired outputs about the testing samples. As can be seen, all testing data are classified to the correct category. The overall recognition rate of the proposed method achieves to 100%.

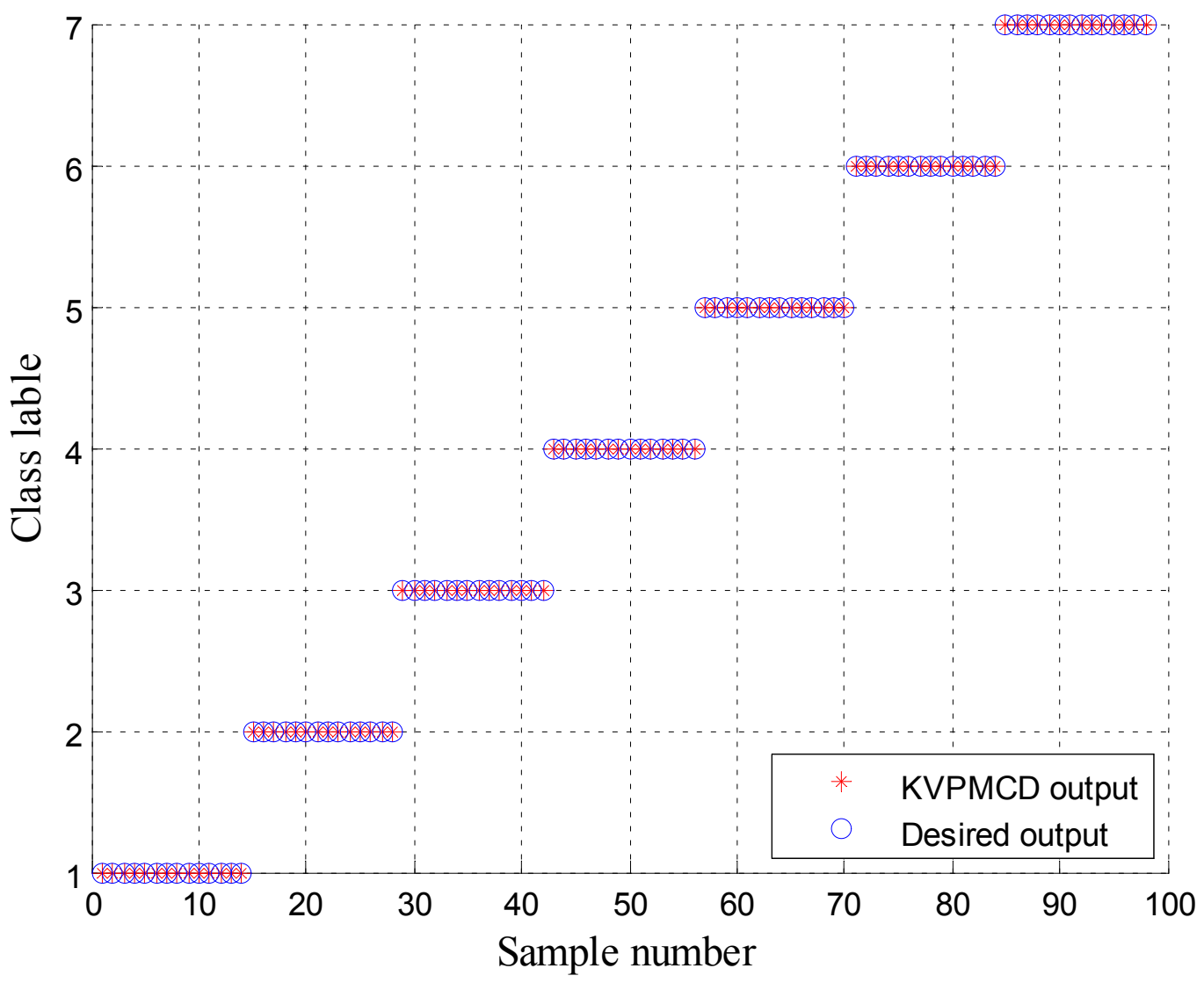

Fig.10 Classification results of the proposed method

To illustrate the optimization of SRCMFE in extracting features, and then the features of each state signal is extracted by MFE is shown in Fig.11. It can be seen from Fig. 11 that, compared to SRCMFE, the first seven features of the MFE are relatively close and the features of the normal state is submerged in the features with faulty states. And Fig.12 shows comparison of standard deviation between SRCMFE and MFE. It can be seen from the Fig.12 that the standard deviation of MFE is almost larger than the standard deviation of SRCMFE at all scales. This shows that the features obtained with SRCMFE have advantages over consistency and stability compared to the features obtained by MFE.

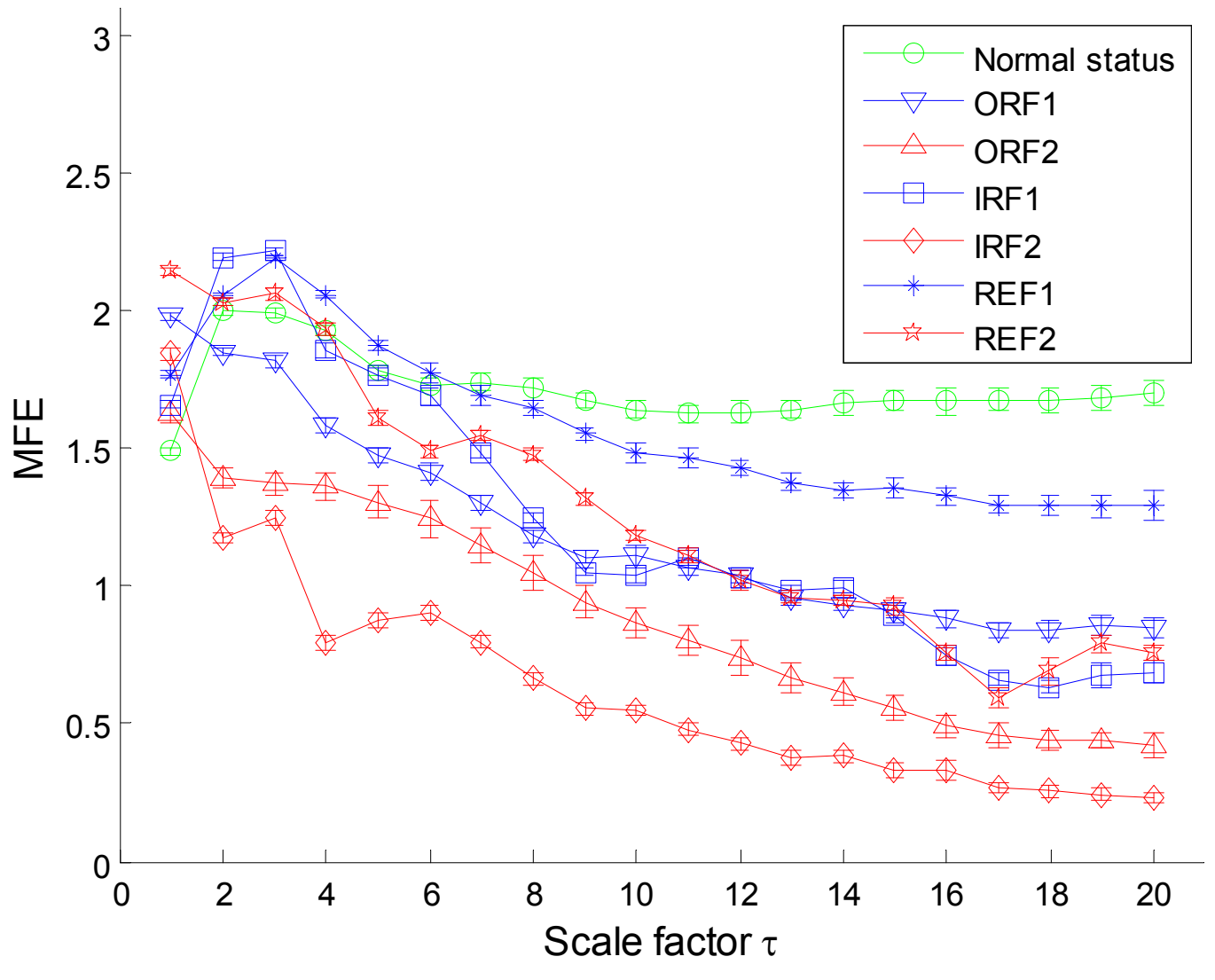

Fig.11 MFE over 20 scales of the corresponding vibration signals

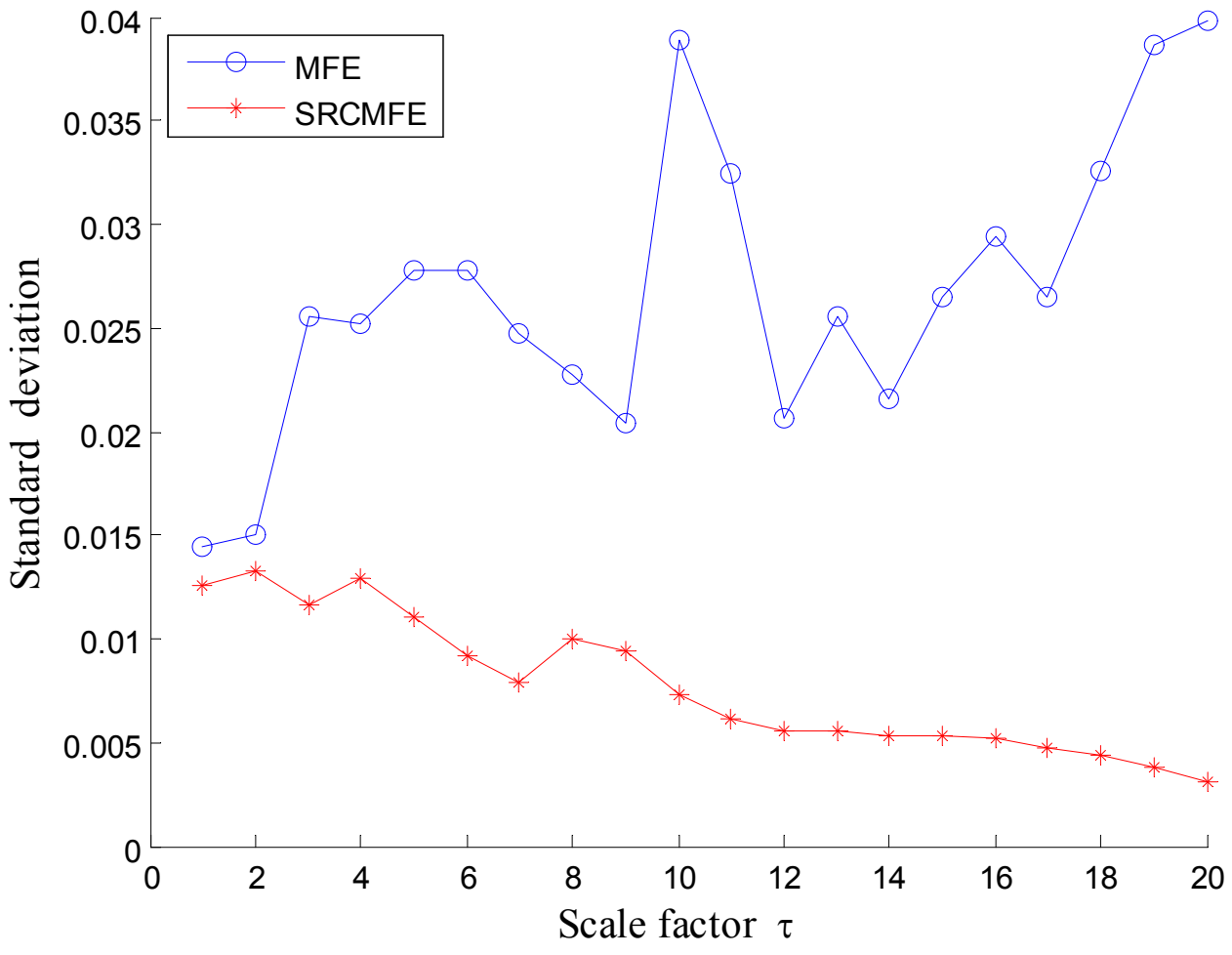

(a) OR1

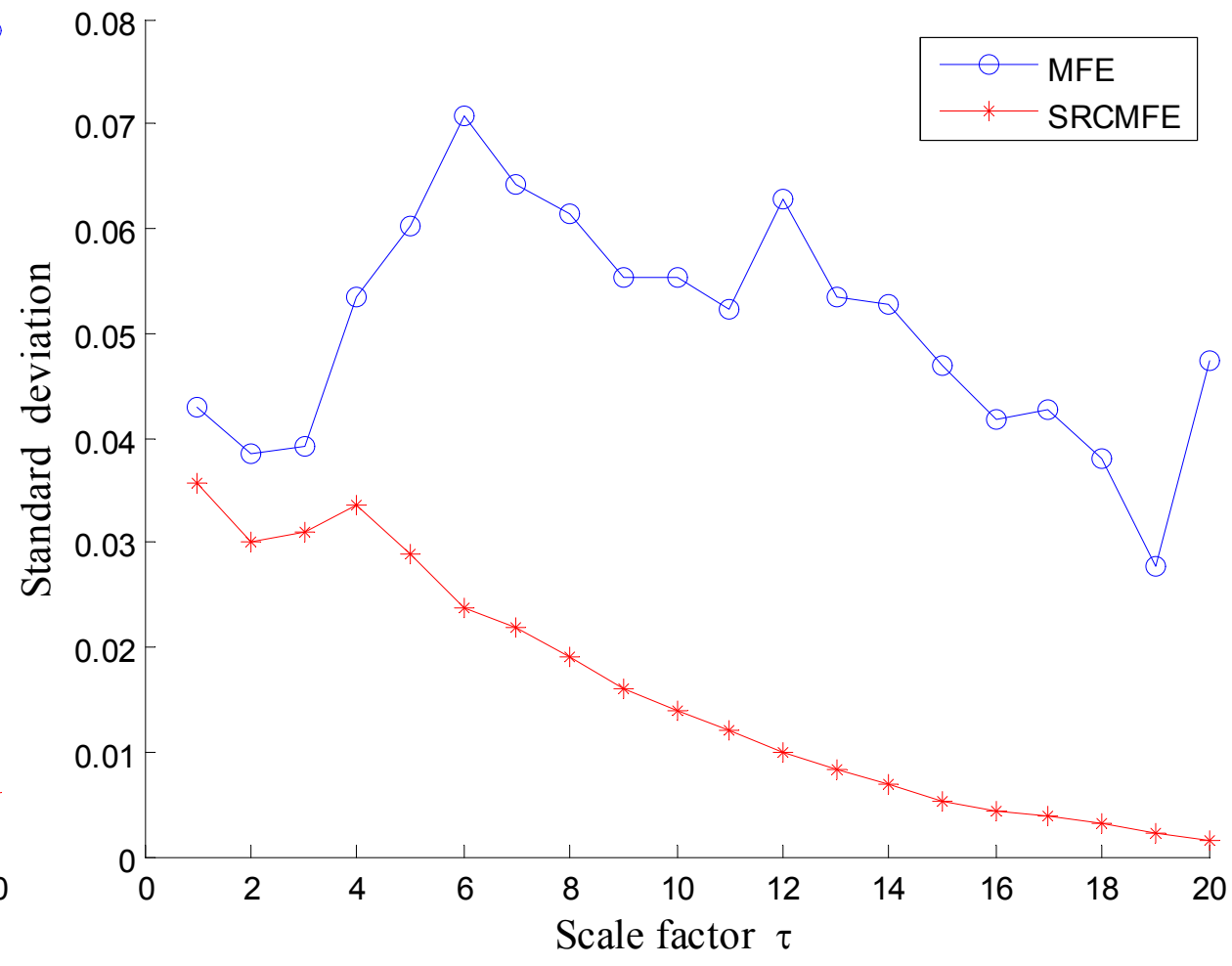

(b) OR2

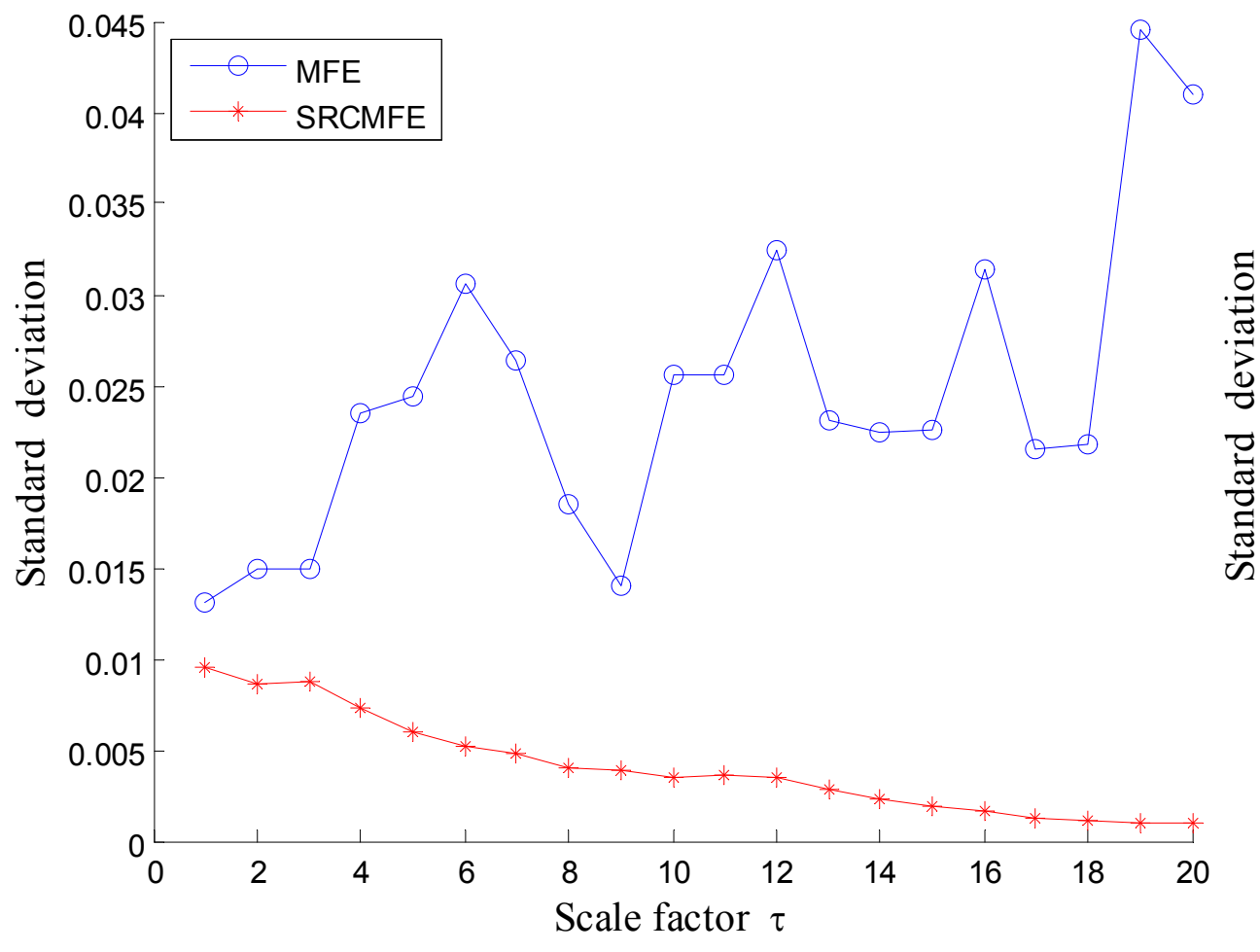

(c) IR1

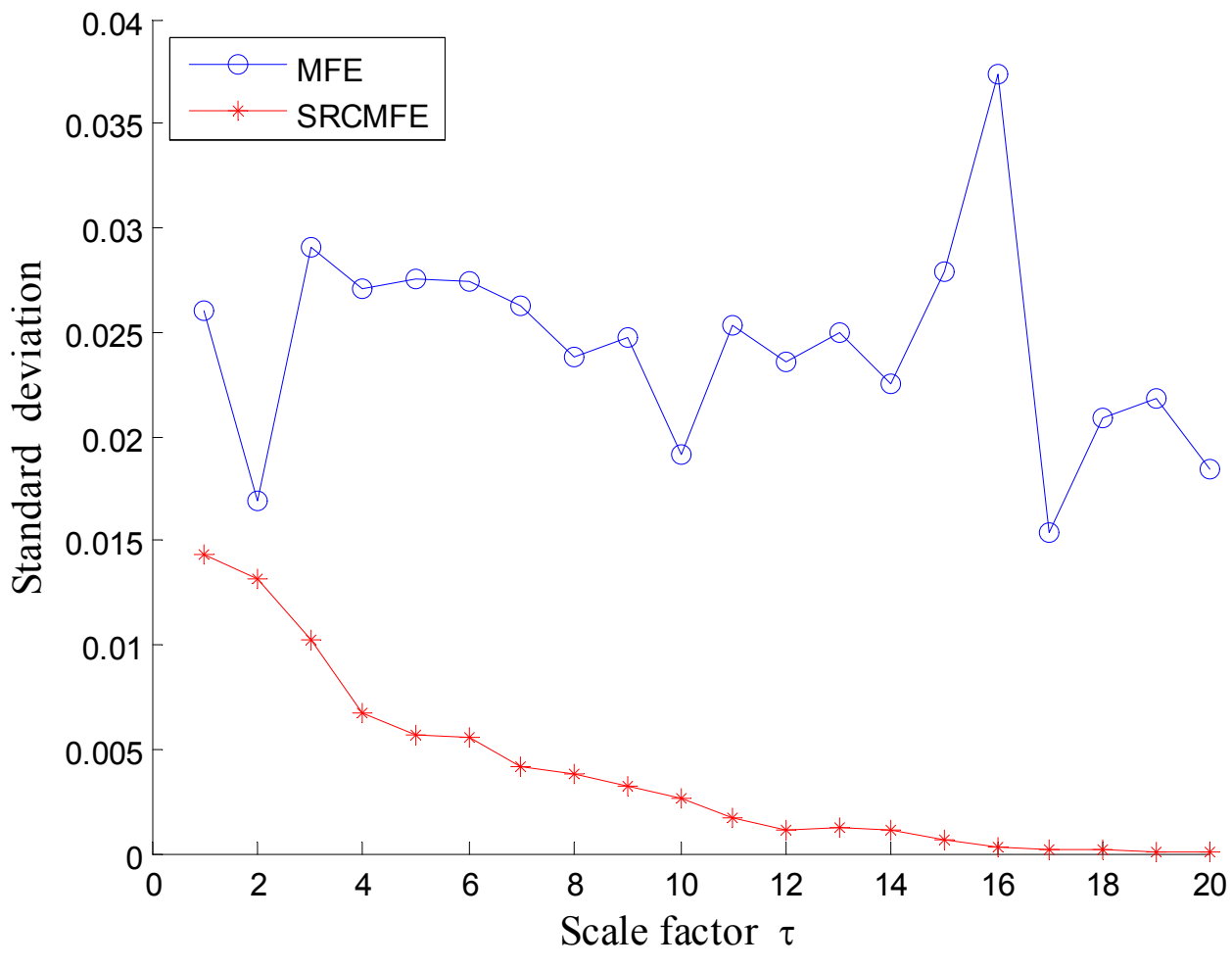

(d) IR2

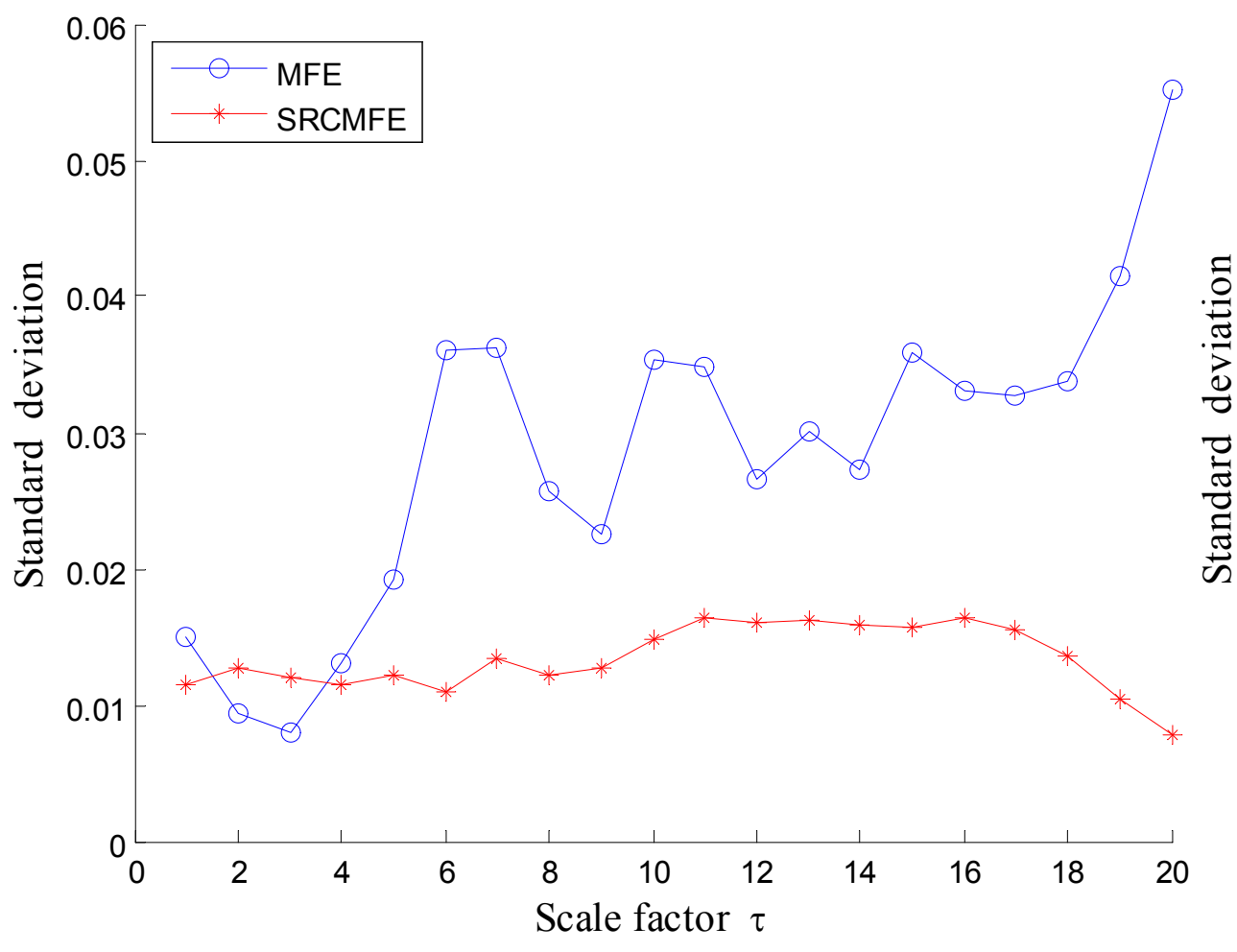

(e) REF 1

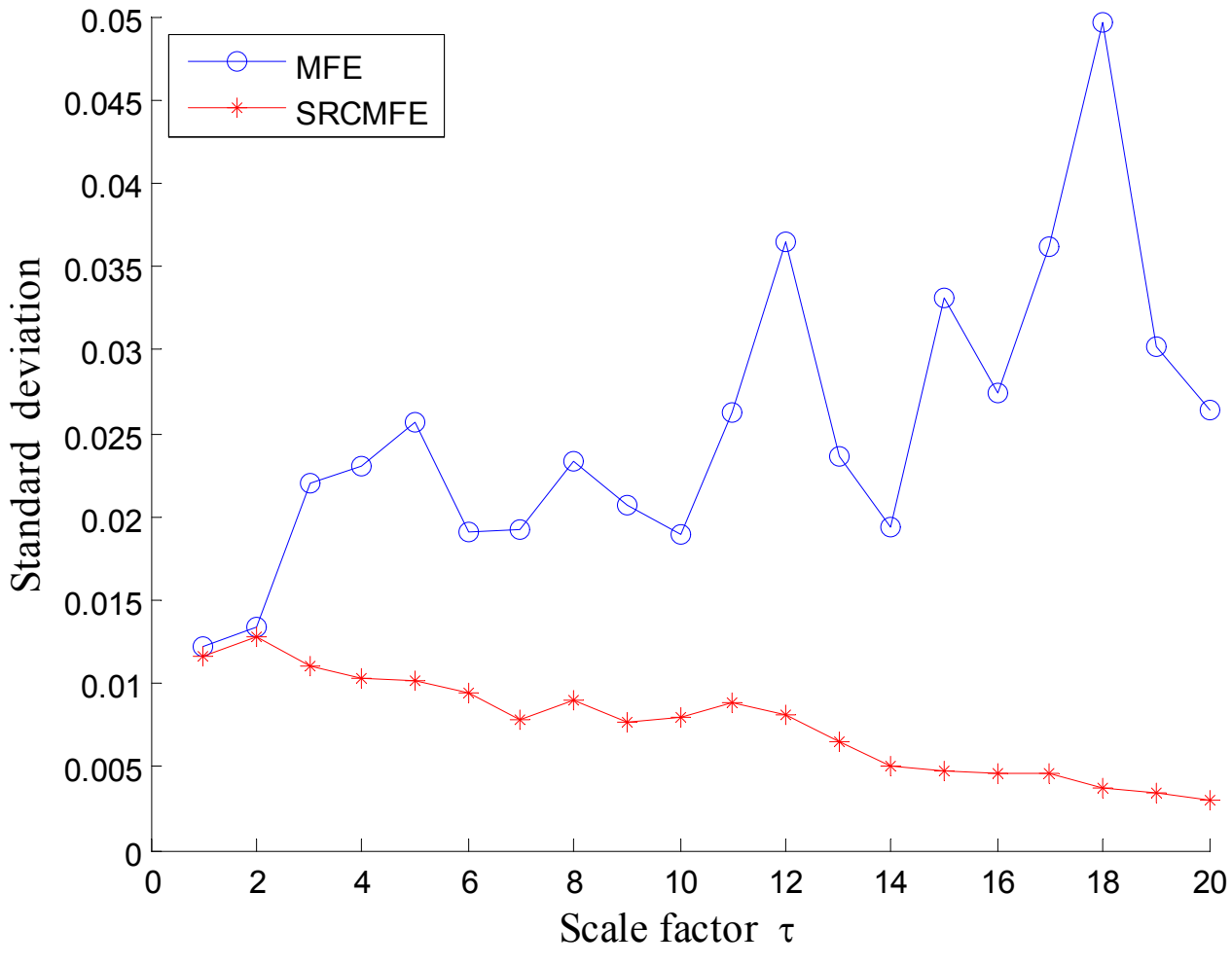

(f) REF2

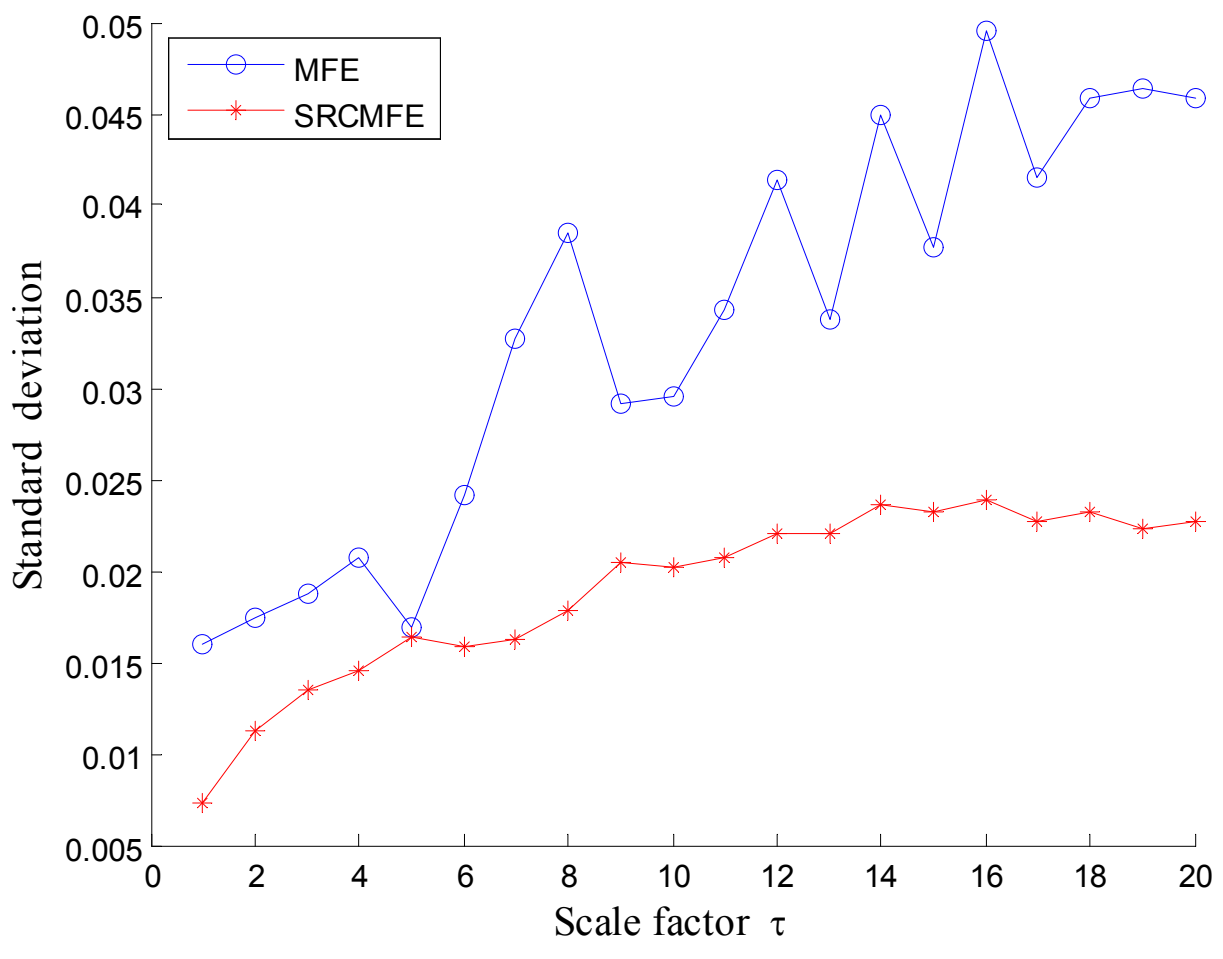

(g) Normal status

Fig.12 Comparison of standard deviation between SRCMFE and MFE

To verify superiority of the dimensionality reduction method used in this paper by using t-SNE algorithm. LDA are employed to dimension reduction respectively after SRCMFE are obtained. As shown in Fig.13. It is clearly seen that normal rolling bearing, inner ring slight fault bearing and outer ring slight fault bearing cluster together, and they have overlapping phenomenon from the Fig.13 (a) and Fig.13 (b). In addition, the contingency tables of classification results of LDA and t-SNE are shown inTables2 for comparisons. It can be seen from the table 2 that the application of t-SNE algorithm able to enhance the accuracy of rolling bearing fault recognition. Therefore, the method using t-SNE is superior to that of using LDA .

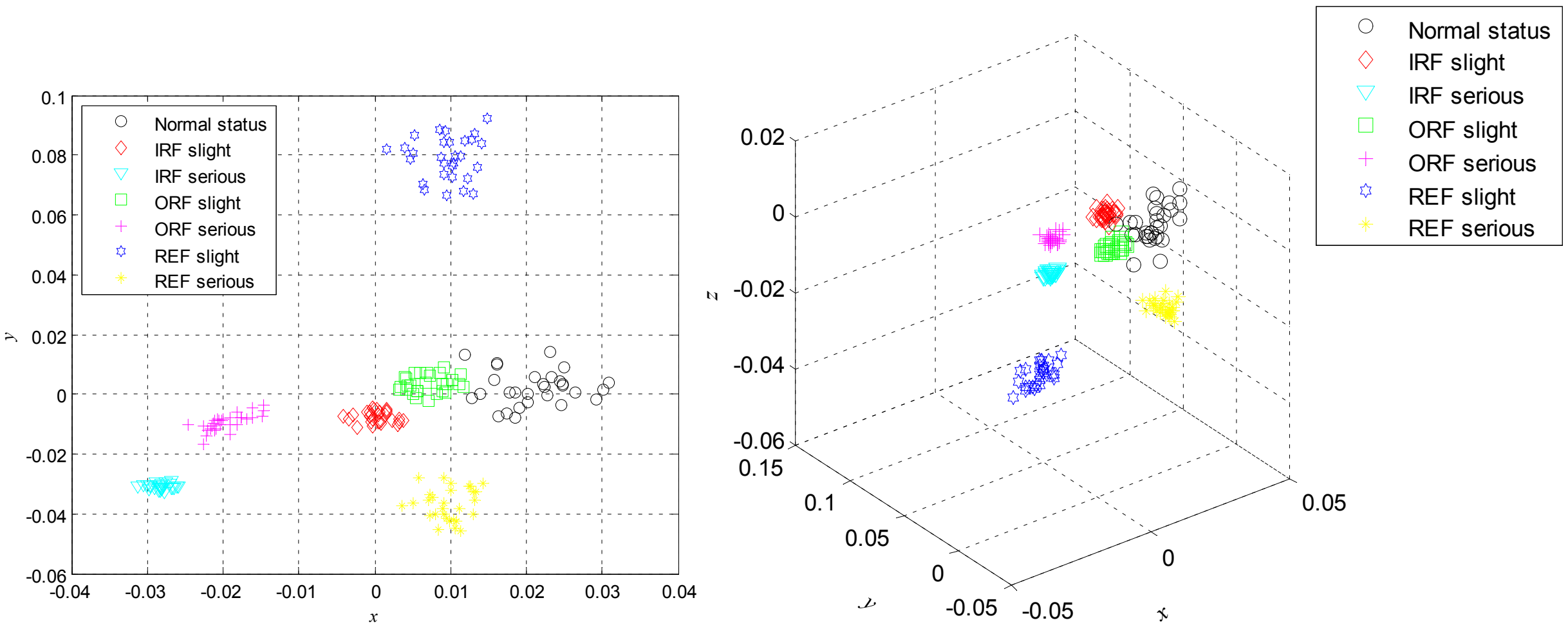

(a) Two-dimensional histogram (b) Three-dimensional histogram

Fig.13 Two-dimensional histogram and three-dimensional histogram by using LDA algorithm (a) & (b)

Table 2

Contingency table of classification results with FS evaluations (datasets for VPMCD training the same as proposed approach)

| Actual | Recognition accuracy |
|---|---|

| class | Normal (%) | Inner race slight fault (%) | Outer race slight fault (%) | Ball element slight fault (%) |
|---|---|---|---|---|
| Recognition accuracy after using LDA | | | | |
| VPMCD | 78.57 | 100 | 100 | 100 |
| Recognition accuracy after using t-SNE | | | | |
| VPMCD | 100 | 100 | 100 | 100 |

| Actual class | Recognition accuracy | | | |
|---|---|---|---|---|
| | Inner race serious fault (%) | Outer race serious fault (%) | Ball element serious fault (%) | Overall recognition rate (%) |
| Recognition accuracy after using LDA | | | | |
| VPMCD | 100 | 100 | 100 | 96.9388 |
| Recognition accuracy after using t-SNE | | | | |
| VPMCD | 100 | 100 | 100 | 100 |

## 6. Conclusions

A novel nonlinear dynamic method termed Sigmoid-based RCMFE is introduced in this paper for measuring the complexity of time series. The SRCMFE is compared with the often used method MFE by analyzing two artificial signals. Since the vibration signals of rolling bearing are nonlinear and non-stationary, as a nonlinear dynamic method, the SRCMFE is very suitable to extract the nonlinear fault features from vibration signals. Based on the SRCMFE, t-SNE for feature selection and VPMCD for fault classification, a new fault diagnosis approach for rolling bearing is proposed and compared with the existing MFE based fault diagnosis methods. Also the t-SNE manifold learning algorithm is employed to feature dimension reduction and is contrasted with the most common used LDA method and the results have indicated the advantages of t-SNE algorithm in dimension reduction. In addition, the novel classification method VPMCD is introduced to achieve rolling bearing fault diagnosis automatically.

**Acknowledgments**

This work was supported by National Key Technologies Research ＆ Development Program of China (No. 2017YFC0805103) and the National Natural Science Foundation of China (Grant No. 51505002) and the Key Program of National Natural Science Foundation of Educational Commission of Anhui Province of China (Grant No. KJ2015A080). Anhui University of Technology Graduate Innovation Research Foundation (Grant No. 2016062).